\documentclass[a4paper]{jpconf}
\usepackage{amsmath,amsfonts,amssymb,amsthm,amsbsy}
\usepackage{diagrams}
\diagramstyle[small,scriptlabels,
 midshaft,nohug]
\newarrow{TeXto}---->
\newarrow{Miss}....>
\newarrow{BothWays}<--->
\newtheorem*{theor}{Theorem}
\newtheorem*{cor}{Corollary}
\theoremstyle{definition}
\newtheorem{prb}{Open problem}[section]
\theoremstyle{remark}
\newtheorem*{example}{Example}
\newcommand{\by}[1]{\textit{{#1}}}
\newcommand{\jour}[1]{\textit{{#1}}}
\newcommand{\vol}[1]{\textbf{{#1}}}
\newcommand{\book}[1]{\textit{{#1}}}
\begin{document}
\title{Homological evolutionary vector fields\\ in
Korteweg\/--\/de Vries, Liouville, Maxwell,
and\\ several other models}

\author{Arthemy V Kiselev}

\address{Johann Bernoulli Institute for Mathematics and Computer Science, University of Groningen,
P.O.~Box 407, 9700~AK Groningen, The~Netherlands}

\ead{A.V.Kiselev@rug.nl}

\begin{abstract}
We review the construction of homological evolutionary vector fields on infinite jet spaces and partial differential equations. We describe the applications of this concept in three tightly inter\/-\/related domains:
the variational Poisson formalism 
(e.g., for equations of Korteweg\/--\/de Vries type),
geometry of Liouville\/-\/type hyperbolic systems 
(including the 2D~Toda chains),
and Euler\/--\/Lagrange gauge theories
(such as the Yang\/--\/Mills theories, gravity, 
or the Poisson sigma\/-\/models). 
Also, we formulate several open problems.
\end{abstract}

\section*{Introduction}
We present a systematic yet very basic review of the 
construction of homological evolutionary vector fields on the infinite jet spaces and on several natural classes of partial differential equa\-ti\-ons such as the gauge models. For a long time, this geometric structure developed in parallel in mathematics and physics. Since mid\/-\/70's, it has been being used intensively in theoretical physics: specifically, within the BRST-{} or BV\/-\/technique for the quantization of gauge\/-\/invariant systems~\cite{BV,BRST,GitmanTyutin,HenneauxTeitelboim,SchwarzBV}. In mathematics, the concept first stemmed over smooth manifolds in the problem of the homological vector field realizations of Lie algebroids~\cite{Vaintrob}, which encompass Lie algebras and vector bundles and are best known in the framework of symplectic and Poisson geometries. Although the notion of Lie algebroids in the geometry of manifolds had been known for a long time~\cite{Mackenzie,Pradines1967}, the proper generalization of this omnipresent structure to the geometry of jet bundles (c.f.~\cite{KumperaSpencer}) appeared in~\cite{Galli10}. In principle, it \emph{could} be contemporary to the discovery of the BRST\/-\/technique (\cite{BRST} vs~\cite{KumperaSpencer}). In the meantime, a steady 
progress in the geometry of jet spaces~\cite{Olver,Vin1984} created the platform for important applications of the future  
concept. Let us name only a few of them:
\begin{itemize}
\item the variational Poisson dynamics for the KdV\/-\/type systems~\cite{DSViniti84,GelfandDorfman}, also~\cite{BelavinSbornik};
\item modern revisions of the theory of Liouville\/-\/type hyperbolic equations~\cite{DSViniti84,SokolovUMN},
in particular, the 2D~Toda chains associated with the Dynkin diagrams~\cite{LeznovSaveliev1979};
\item the cohomological approach to gauge fields~\cite{BarnichBrandtHenneaux,BarnichBrandtHenneauxPhysRep}
and (quantum) Poisson theory~\cite{BelavinSbornik}.
\end{itemize}
In this paper, we recall the jet\/-\/bundle geometry that stands behind the structures~$\boldsymbol{Q}^2=0$. It allows us to address, from a unified standpoint~\cite{McCloud,SchwarzBV}, 
many relevant physical models ranging from the Korteweg\/--\/de Vries equation and its Poisson structures~\cite{Magri} to the Yang\/--\/Mills theories, gravity, and Poisson sigma\/-\/models~\cite{AKZS,CattaneoFelder2000}.

In section~\ref{SecInvolutive} we first consider the evolutionary vector fields
$\partial^{(\boldsymbol{q})}_\varphi=\sum_{|\sigma|\geq0}
\tfrac{\mathrm{d}^{|\sigma|}}{\mathrm{d}\boldsymbol{x}^\sigma} (\varphi)\cdot \partial/\partial\boldsymbol{q}_\sigma$ whose generating sections $\varphi\bigl(\boldsymbol{x},[\boldsymbol{q}]\bigr)$ belong to the images of matrix linear 
operators $A_1$,\ $\ldots$,\ $A_N$ in total derivatives 
(and 
those derivatives are equal to $\tfrac{\mathrm{d}}{\mathrm{d}x^i}=\tfrac{\partial}{\partial x^i}+\sum_{|\sigma|\geq0}\boldsymbol{q}_{\sigma+1_i}\cdot\partial/\partial\boldsymbol{q}_\sigma$). The three above\/-\/mentioned classes of geometries are regular sources of collections of such operators. We impose the Berends\/--\/Burgers\/--\/van~Dam hypothesis~\cite{BBvD1985}: namely, we let the images of the operators be subject to the collective commutation closure. This involutive setup gives rise to the bi\/-\/differential structure constants~\cite{FradkinVasiliev} and bi\/-\/differential analogs of Christoffel symbols~\cite{Christoffel}.
For any number~$N>1$ of the operators $A_1$,\ $\ldots$,\ $A_N$, we reduce the setup at hand to the variational Lie algebroids~\cite{Galli10} with one variational anchor. 
Having endowed the new bundle geometry 
with the Cartan connection, we reverse the parity of the fibres in it by brute force and then construct the odd, parity\/-\/reversing evolutionary vector field~$\boldsymbol{Q}$ that encodes the entire initial geometry
via the homological equality~$\boldsymbol{Q}^2=0$.

In section~\ref{SecExamples} we describe three natural examples of such geometries: the variational Poisson algebroids~\cite{JKGolovko2008,Galli10}, 2D~Toda chains viewed as variational Lie algebroids~\cite{Galli10,SymToda}, and gauge algebroids~\cite{BarnichBialowieza,Christoffel}.

First, in section~\ref{SecHam} we recall the construction of necessary superbundles using the 
notion of the variational cotangent bundle~\cite{KuperCotangent} and derive the representation via~$\boldsymbol{Q}^2=0$ for the variational Poisson structures~\cite{GelfandDorfman}. As usual, the renowned KdV equation~\cite{Magri} offers us the minimally possible nontrivial illustration. The odd vector fields~$\boldsymbol{Q}$ are themselves Hamiltonian and we write the corresponding $W$-charges explicitly.

By exploiting the profound relation between the variational Poisson geometry and 2D~Toda chains~\cite{DSViniti84,TMPhGallipoli,SymToda,SokolovUMN}, we obtain the evolutionary fields~$\boldsymbol{Q}$ for these models of Liouville type~\cite{YamilovShabat}. We emphasize that the differentials~$\boldsymbol{Q}$ and their cohomology carry much more information than the ordinary Chevalley\/--\/Eilenberg differentials for the semi\/-\/simple Lie algebras at hand. 
Being obtained from the gauge\/-\/invariant Yang\/--\/Mills equations under a symmetry reduction~\cite{LeznovSaveliev1979}, the 2D~Toda chains represent the vast class of nonlinear Euler\/--\/Lagrange systems of Liouville type~\cite{Saveliev1992,YamilovShabat,SokolovUMN}. This class is very interesting by itself; moreover, the construction of the homological vector fields~$\boldsymbol{Q}$ in section~\ref{SecToda} remains valid uniformly for all such equations.

Finally, in section~\ref{SecGauge} we expose the true 
geometric nature of the BRST\/-\/differentials~\cite{BRST} or the `longitudinal' components of the BV\/-\/differentials~\cite{BV} for gauge systems. In agreement with the Second Noether Theorem, the differential operators~$A_1$,\ $\ldots$,\ $A_N$ which we study here emerge from the Noether relations
between the equations of motion; we explain why the same approach of $\boldsymbol{Q}^2=0$ can successfully grasp more subtle geometries without any further modifications.
The gauge parameters (that is, the arguments of the operators~$A_i$), being the parity\/-\/reversed \emph{neighbours} of the
\emph{ghosts}, together with the \emph{antifields} (which will have appeared before in section~\ref{SecHam}) and the newly introduced \emph{antighosts} constitute the classical BV\/-\/zoo. Similarly to the variational Poisson case, the evolutionary BV\/-\/differentials~$\boldsymbol{D}$, which incorporate the odd vector fields~$\boldsymbol{Q}$, are then determined by the Schouten bracket of the BV\/-\/charges with the inhabitants of the~zoo.

In the appendix, we report the results of a direct search for the linear differential operators with involutive images; their coordinate\/-\/free understanding as the variational anchors in variational Lie algebroids is yet to be achieved.

We conclude that the presence of the evolutionary differentials~$\boldsymbol{Q}$ is not specific to either the variational Poisson or gauge systems alone: indeed, it is an immanent property of a much wider class of models. The study of the arising cohomology theories not only sheds new light on each system or structure in particular but also reveals far\/-\/reaching links between them.

\section{Preliminaries: Notation and conventions}
\rightline{\parbox{9.5cm}{\strut\textsf{This towel may not serve
as a shelter during the hurricane.}\\
\mbox{ }\hfill(Liability disclaimer)}}

\medskip
The geometry of jet fibre bundles is an established and well developed domain~\cite{BarnichBrandtHenneaux,ClassSym,
McCloud,Olver,Saunders,Vin1984}. Any deviations between the notations from different sources are easily tractable; we follow the notation of~\cite{Galli10,Christoffel}
and adopt Dirac's convention on bra-{} covectors~$\langle\,\mid$ and {}-ket vectors~$\mid\,\rangle$, 
fixing the orientation~$\delta\boldsymbol{p}\wedge\delta\boldsymbol{q}$ on the cotangent bundles.

By default, we pronounce the standard incantation about the ground field~$\mathbb{R}$, the infinite smoothness of all mappings at hand, the locality along the oriented base~$\varSigma^n$ of the bundles (whence emerge the jets of sections of \emph{vector} bundles $\pi\colon E^{n+m}\to\varSigma^n$ instead
of the jet spaces $J^\infty(\varSigma^n\to M^m)$ for maps of manifolds), and the locality with respect to the jet order (whence the filtration in $C^\infty(J^\infty(\pi))$). We assume that all total differential operators are local, i.e., polynomial in the total derivatives. By convention, the symmetries are infinitesimal and we identify the evolutionary vector fields~$\mid\partial_\varphi\rangle$ with their generating sections~$\mid\varphi\rangle$. For the sake of transparency, we consider the purely even, non\/-\/graded initial geometry of commutative vector spaces, manifolds, and bundles; the odd objects which we deal with appear later, through the explicit use of the parity reversion~$\Pi$.

Our reasonings are independent of the choice of local coordinates. We denote the independent variables by $\boldsymbol{x}\in\varSigma^n$, whereas the notation $\boldsymbol{q}$ runs through all the dependent variables such as the gauge fields (if any in the model under study) at the points of space\/-\/time or say, the realizations of the ``string''~$\varSigma^n$ in the space\/-\/time~$M^{3,1}$. The notation is multi\/-\/faceted: in each class of examples (see section~\ref{SecExamples}), we shall recall the common
specific notation so that, on one hand, an object in this review may acquire several interpretations in different contexts, while on the other hand, we shall of course meet with the habitual ghosts, antighosts, and antifields.

We work with the infinite jet bundles from the very beginning. Let us remember that many elementary constructions on usual manifolds
--first of all, the Leibniz rule in the commutator~$[\,,\,]$ and the Schouten bracket~$[\![\,,\,]\!]$ or in the Poisson brackets under the multiplication of the arguments by smooth functions-- either entirely disappear over the jet spaces or (seldom) survive after a careful consideration and proper amendments. Furthermore, we deal with the variational Poisson structures~\cite{GelfandDorfman,Lstar} over the spaces $J^\infty(\pi)$ of jets of sections for vector bundles $\pi\colon E^{n+m}\to\varSigma^n$ rather than suppose that the target space $M^m$ for $J^\infty(\varSigma^n\to M^m)$ alone is a Poisson manifold (c.f.\ \cite{AKZS,CattaneoFelder2000,FulpLadaStasheffSrni}). 
The variational Poisson theory we exploit does not amount to the canonical formalism which one applies to a Lagrangian system after taking its Legendre transformation (except for
section~\ref{SecToda} where we do use that approach as well),
c.f.~\cite{BarnichHenneauxBVPoisson,Nutku2000}.

For completeness, we explicitly endow the horizontal infinite jet bundles~\cite{Lstar} over the spaces~$J^\infty(\pi)$ with the Cartan connection. When introducing the modules of sections of the horizontal bundles such as the spaces of the gauge parameters, never we attempt to treat these sections 
as the ``collections of functions'' or use any other, equally weak interpretation of their structure because it does not grasp properly the geometry of the building blocks --- instead, this shadows it with irrelevant fragmentations.
Indeed, let us remember that, under a differential change of the coordinates~$\boldsymbol{x}$ and~$\boldsymbol{q}$, the genuine reparametrization rules for those sections are imperatively prescribed by their geometric nature (see~\cite{Galli10,SymToda} for detail and examples). We thus ought to be careful with the $C^\infty(J^\infty(\pi))$-\/module structure for the spaces of such sections; this is specific in particular to the variational anchors in variational Lie algebroids (see section~\ref{SecInvolutive} and~\cite{Galli10}) which are neither the anchors nor Lie algebroids for usual manifolds, respectively, unless the base manifold~$\varSigma^n$ shows itself in space\/-\/time as no more than a point, or unless all gauge fields are uniformly constant. This also means that we deal \textit{ab initio} with the field\/-\/dependent gauge parameters $\boldsymbol{p}\bigl(\boldsymbol{x},[\boldsymbol{q}]\bigr)$; moreover, their rigorous introduction (see section~\ref{SecGauge}) demonstrates that the gauge models must be particularly restrictive in order for these sections~$\boldsymbol{p}$ to not depend on a part of the variables that encode the points of~$J^\infty(\pi)$.

In sections~\ref{SecInvolutive} and~\ref{SecHam} we work on the empty jet spaces; later, in sections~\ref{SecToda} and \ref{SecGauge}, we do on\/-\/shell. 
We assume the off\/-\/shell validity of the Berends\/--\/Burgers\/--\/van~Dam hypothesis stating the collective commutation closure of the evolutionary fields~$\partial^{(\boldsymbol{q})}_{A_i(\cdot)}$ whose generating sections~$A_i(\cdot)$ belong to the images of linear differential operators~$A_i$. This may require us to 
quotient out the trivial symmetries that vanish on\/-\/shell.

We recall that the Euler\/--\/Lagrange systems $\mathcal{E}_{\text{EL}}=\bigl\{ F_i\mathrel{{:}{=}}\delta S/\delta q^i=0\bigr\}$ contain as many equations as there are unknowns 
in~it. (In contrast with~\cite{FulpLadaStasheffSH}, we do not assume that the density of the action functional~$S$ is a differential polynomial: indeed, see section~\ref{SecToda}.)
We point out also that \emph{usually} the equations, either in evolutionary systems or in the Euler\/--\/Lagrange systems, are enumerated (more precisely, labelled) by the respective unknowns~$q^i$ which explicitly occur in the left\/-\/hand sides. However, we emphasize that the admissible reparametrizations of the fields~$\boldsymbol{q}$ and of the equations~$\boldsymbol{F}=0$ are entirely unrelated. This produces its due effect on the transcription of the infinitesimal symmetries $\dot{\boldsymbol{q}}=\varphi\bigl(\boldsymbol{x},[\boldsymbol{q}]\bigr)$ in the former case and, in the latter, on the objects that lie in or are dual to the horizontal module $P_0\ni F_i$ of the equations, e.g., on the antifields~$\boldsymbol{q}^\dagger$ (thus, more appropriately denoted by~$\boldsymbol{F}^\dagger$), the generating sections~$\psi$ of conservation laws (for the Euler\/--\/Lagrange systems, emerging from the Noether symmetries~$\varphi_{\mathcal{L}}$ by the First Noether Theorem), or the objects which obey the Second Noether Theorem such as the ghosts~$\boldsymbol{\gamma}$ and the antighosts~$\boldsymbol{\gamma}^\dagger$. The 2D~Toda chains in section~\ref{SecToda} offer a perfect example of such discorrelation between~$\varphi_{\mathcal{L}}$ and~$\psi$, which is brought in by force due to a purely \ae s\-the\-tic tradition of writing hyperbolic systems with their symbols cast into diagonal shape. 
This requires a bit of attention (see~\cite{TMPhGallipoli,SymToda}): the Noether map $\varphi_{\mathcal{L}}\to\psi$ is not always the identity.

In this paper, we understand as the \emph{Noether maps} not only the linear differential operators $\widehat{P_0}\ni\psi\mapsto\varphi 
\in\text{sym}\,\mathcal{E}$ but, for~$i\geq0$, the operators
$\widehat{P_i}\to\text{sym}\,\mathcal{E}$ at each $i$th generation of the \emph{syzygies}
$\boldsymbol{\Phi}_i[\boldsymbol{\Phi}_{i-1}]\in P_i$, i.e., the equations
$\mathcal{E}=\{\boldsymbol{F}=0\mid F_j\in P_0\}$ at~$i=0$,
Noether's relations $\boldsymbol{\Phi}_1[\boldsymbol{F}]\equiv0$ at~$i=1$, etc.; the notation~$\widehat{P_i}$ corresponds to the modules of sections dual to~$P_i$. In this sense, the Hamiltonian operators $A\colon\widehat{P_0}\to\text{sym}\,\mathcal{E}$ and the generators $A_i\colon\widehat{P_1}\to\text{sym}\,\mathcal{E}$ of gauge symmetries are Noether operators.

On the same footing, we note that the arguments~$\boldsymbol{p}_i$ of the linear differential operators~$A_i$ which determine the evolutionary vector fields~$\partial^{(\boldsymbol{q})}_{A_i(\boldsymbol{p}_i)}$ do possess an ambiguity. This is due to several independent reasons and occurs not only because of the on\/-\/shell equivalence:
for a given equation~$\mathcal{E}$, we let $\boldsymbol{p}'_i\sim\boldsymbol{p}''_i$ whenever $\boldsymbol{p}'_i-\boldsymbol{p}''_i\approx0$ on~$\mathcal{E}$. But indeed, the kernels~$\ker A_i$ in reducible gauge theories carry the freedom as large as the presence of the free functional parameters in their description if the operator equations $A_i\circ\nabla_i=0$ admit nontrivial solutions~$\nabla_i\neq0$. Independently, the differential equations $\sum_{i=1}^N A_i\bigl(\boldsymbol{p}_i\bigl(\boldsymbol{x},
[\boldsymbol{q}(\boldsymbol{x})]\bigr)\bigr)=0$ may have non\/-\/empty non\/-\/linear spaces of solutions~$\boldsymbol{q}(\boldsymbol{x})$. 
This shows that the structure constants~\eref{BiDiffC} (c.f.\ \cite{FradkinVasiliev} and~\cite{Christoffel}), which are bi\/-\/differential operators with respect to the arguments~$\boldsymbol{p}_1$,\ $\ldots$,\ $\boldsymbol{p}_N$ of the operators~$A_1$,\ $\ldots$,\ $A_N$, are in fact equivalence classes of mappings. Because of this, considerable efforts are required to achieve the canonical normalization of the homological evolutionary vector fields on the jet\/-\/bundle extensions of such domains, see~\cite{Galli10} and~\cite{FulpLadaStasheffSH}.

The construction of gauge symmetries, which appear in a class of geometries very particular by themselves, is reviewed in section~\ref{SecGauge}. But let us recall in advance that these symmetries have the form~$\partial^{(\boldsymbol{q})}_{A_i(\boldsymbol{p}_i)}$ with arbitrary $\boldsymbol{p}_i=\boldsymbol{p}_i\bigl(\boldsymbol{x},[\boldsymbol{q}]\bigr)$. In this context, the class of hyperbolic partial differential equations $\boldsymbol{q}_{xy}=\boldsymbol{f}(x,y;q^i,q^j_x,q^k_y)$ that admit arbitrary functions~$p(x)$ and $\overline{p}(y)$ in their symmetries is \emph{close~to
but different} from the renowned class of Liouville\/-\/type hyperbolic quasilinear systems (see section~\ref{SecToda} or~\cite{SokolovUMN,SokStar} and~\cite{TMPhGallipoli,SymToda}).

Summarizing, we always highlight the geometric nature of the structures at hand and describe the links between them, as carefully as possible. The literature which is available on the rapidly developing topic of this review is so vast that,
instead of the botany of new remarkable advancements,
we now do the sociology of repeatedly practised oversights.

\section{Basic concept: Involutive distributions of operator\/-\/valued evolutionary 
vector fields}\label{SecInvolutive}
Let $\pi$ and~$\xi_i$, $i=1,\ldots,N$, be vector bundles over the base manifold~$\varSigma^n\ni\boldsymbol{x}$ and denote by $\pi_\infty\colon J^\infty(\pi)\to\varSigma^n$ and $\xi_{i,\infty}\colon J^\infty(\xi_i)\to\varSigma^n$ the respective projections (the forgetful maps) to the base of the spaces of infinite jets of sections for these bundles. Consider the \emph{horizontal modules} $\Gamma\bigl(\pi_\infty^*(\xi_i)\bigr)$ of sections of the induced bundles~$\pi_\infty^*(\xi_i)$ and denote their elements by~$\boldsymbol{p}_i\bigl(\boldsymbol{x},[\boldsymbol{q}]\bigr)$. The main example of such modules is $\varkappa(\pi)\mathrel{{:}{=}}\Gamma\bigl(\pi_\infty^*(\pi)\bigr)\ni\varphi$, which is the $C^\infty(J^\infty(\pi))$-\/module of the sections of evolutionary derivations $\partial^{(\boldsymbol{q})}_{\varphi}=\varphi\,\tfrac{\partial}{\partial\boldsymbol{q}}+ \tfrac{\mathrm{d}}{\mathrm{d}\boldsymbol{x}}(\varphi)\,\tfrac{\partial}{\partial\boldsymbol{q}_{\boldsymbol{x}}}+\cdots$ on~$J^\infty(\pi)$; we shall also use the 
modules~$\overline{\Lambda}^n(\pi)$ of the highest horizontal forms\footnote{The $C^\infty(J^\infty(\pi))$-\/module structure for the spaces of Hamiltonians~$\mathcal{H}$ or Lagrangians~$\mathcal{L}$ is practically useless because it does not pass --via any reasonable Leibniz rule-- through the variational Poisson or Schouten brackets.}
$\mathcal{L}=L\cdot\pi_\infty^*(\mathrm{d}\text{vol}\,
(\varSigma^n)$
and the modules~$\widehat{\varkappa(\pi)}$ of the \emph{variational covectors}~$\psi$ which are dual\footnote{The agreed choice of the coupling implicitly prescribes that we fix the metric and the volume form on the base~$\varSigma^n$.
Let us add that in this paper we never focus on the concrete choice of that metric and also we do not pay any \emph{particular} attention to the upper or lower location of the indices.}
to the sections~$\varphi\in\varkappa(\pi)$ with respect to the coupling $\langle\,,\,\rangle$ that takes values in~$\overline{\Lambda}^n(\pi)$. (Generally speaking, the construction of the bundles~$\xi_i$ can be very involved even in the regular and natural classes of geometries, see section~\ref{SecToda} and~\cite{SymToda,Galli10}.) Not accenting it at all times, we shall project to the $n$th horizontal cohomology spaces~$\overline{H}^n(\pi)$ that are determined by the lifting~$\overline{\mathrm{d}}$ onto~$J^\infty(\pi)$ of the de Rham differential on~$\varSigma^n$. By this we gain the right to integrate by parts; we denote by $\int\omega$ or by $[\omega]$ the equivalence classes of 
differential forms~$\omega$.

Suppose that $A_i\colon\Gamma\bigl(\pi_\infty^*(\xi_i)\bigr)\to\varkappa(\pi)$ are linear total differential operators
that take values in the space of evolutionary vector fields for each $i=1,\ldots,N$. That is, the domains of the~$N$ operators may be different but the target space is common to them all. The Hamiltonian operators $A\colon\widehat{\varkappa(\pi)}\to\varkappa(\pi)$ on empty jet spaces~$J^\infty(\pi)$ are an example of such structures~\cite{GelfandDorfman}.

Assume further that the images of the~$N$ operators~$A_i$ are subject to the collective commutation closure,
\begin{equation}\label{Involutive}
[\text{im}\,A_i,\text{im}\,A_j]\subseteq\sum\nolimits_{k=1}^N\text{im}\,A_k.
\end{equation}
The sum of the images in the right\/-\/hand side is not direct because in effect they can overlap~\cite{Magri}.
Assumption~\eref{Involutive} gives rise to the bi\/-\/differential structure constants~\cite{FradkinVasiliev} via
\[
\bigl[A_i(\boldsymbol{p}_i),A_j(\boldsymbol{p}_j)\bigr] =
\sum\limits_{k=1}^N A_k\bigl(\mathbf{c}_{ij}^k(\boldsymbol{p}_i,\boldsymbol{p}_j)\bigr);
\]
each $\mathbf{c}_{ij}^k$ is the equivalence class of bi\/-\/differential operators (possibly, with \emph{non}constant, field\/-\/dependent coefficients) with both its arguments and its values taken modulo the kernels of the operators~$A_i$,\ $A_j$ and~$A_k$, respectively. (The kernels can be sufficiently large unless the extra nondegeneracy assumptions are made; for the sake of brevity, we shall not emphasize the presence of such kernels in \emph{all} formulas but rather in few ones, c.f.\ \cite{Galli10}.) 

Using the permutability of the evolutionary vector fields with the total derivatives and taking into account the Leibniz rule that always holds for the evolutionary derivations~$\partial^{(\boldsymbol{q})}_{A_\ell(\boldsymbol{p}_\ell)}$, we obtain the canonical decomposition of the 
bi\/-\/differential structure constants,
\begin{equation}\label{BiDiffC}
\mathbf{c}_{ij}^k(\boldsymbol{p}_i,\boldsymbol{p}_j)=
\partial^{(\boldsymbol{q})}_{A_i(\boldsymbol{p}_i)}(\boldsymbol{p}_j)\cdot\delta_j^k
 -\partial^{(\boldsymbol{q})}_{A_j(\boldsymbol{p}_j)}(\boldsymbol{p}_i)\cdot\delta_i^k
 +\Gamma_{ij}^k(\boldsymbol{p}_i,\boldsymbol{p}_j).
\end{equation}
The symbols~$\Gamma_{ij}^k$ absorb the bi\/-\/linear action of~$\mathbf{c}_{ij}^k$ by total differential operators on the arguments; the notation is justified~\cite{Christoffel} because the terms~$\Gamma_{ij}^k$ do indeed transform as the Christoffel symbols under the reparametrizations $\boldsymbol{p}_i\mapsto g\,\boldsymbol{p}_i$ invoked by a differential change of the jet coordinates (whence the $g$'s are linear
operators in total derivatives). 
We emphasize that generally speaking, the symbols~$\Gamma_{ij}^k$ alone --that is, \emph{without} the first two standard terms in~\eref{BiDiffC}-- do not determine any Lie algebra structure: the true Jacobi identity 
is given by~\eref{JacobiGamma} 
where the standard terms are taken into account properly.
It is likely that the restoration of the standard component in~\eref{BiDiffC} for known bi\/-\/differential symbols~$\Gamma_{ij}^k$ is the cause of a fierce struggle in the modern double field theory and in the theories which are based on Courant\/-\/like brackets. 
We notice further that the bi\/-\/differential Christoffel symbols introduced in~\eref{BiDiffC} must not be confused with the connections $\boldsymbol{\mathcal{A}}=\mathcal{A}_i\,\mathrm{d}x^i$ in principal fibre bundles in gauge theories.
It is interesting to pursue further this emerging parallel between the geometry over the jet bundles and the known connection\/-\/based structures over usual manifolds.

We remark that, in general, $\mathbf{c}_{ii}^k\neq0\pmod{\ker A_k}$ if $k\neq i$, that is, the image of an operator alone may not be involutive even if the operator (in particular, its domain of definition) is well defined regardless of the entire collection~$A_1$,\ $\ldots$,\ $A_N$. Also, we note that for only one operator ($N=1$) the Lie algebra of evolutionary vector fields in its image is, generally, non\/-\/abelian, which must not be confused with the abelian gauge theories (e.g., Maxwell's electrodynamics) corresponding to one\/-\/dimensional Lie groups, which are always commutative.

We finally notice that the nature of the arguments~$\boldsymbol{p}_i$ is firmly fixed by the reparametrization rules $\boldsymbol{p}_i\mapsto g\,\boldsymbol{p}_i$. Therefore, the isolated components of the sections~$\boldsymbol{p}_i$, which one may be tempted to treat as ``functions'' and by this fully neglect their geometry, may not be well\/-\/defined as true \emph{functions}. We repeat that this primitivization discards much of the information about the setup and shadows the geometry which we have addressed so far.

\smallskip
We now reduce, in a regular way, the setup of \emph{many} operators and their individual domains to one operator on the large domain~\cite{Galli10}. Namely, we take the Whitney sum over~$J^\infty(\pi)$ of the induced bundles $\pi_\infty^*(\xi_i)$, which in down\/-\/to\/-\/earth terms means that we compose the new ``tall'' sections
\begin{subequations}\label{BothTallWide}
\begin{align}
\boldsymbol{p}&={}^{\mathrm{t}}\bigl(\boldsymbol{p}_1\mid
\boldsymbol{p}_2\mid\ldots\mid\boldsymbol{p}_N\bigr)\in
\Gamma\Bigl(\mathop{{\bigoplus}_{J^\infty(\pi)}}\limits_{k=1\phantom{MMll}}^{N\phantom{MMll}}\pi_\infty^*(\xi_i)\Bigr)\mathrel{{=}{:}}
\Gamma\Omega\bigl(\xi_\pi\bigr),
\label{TallSection}\\
\intertext{and understand the 
sections~$\boldsymbol{p}$ as the arguments of the new ``wide'' matrix operator}
A&=\bigl(A_1\mid A_2\mid\ldots\mid A_N\bigr).\label{WideOperator}
\end{align}
\end{subequations}
Obviously, property~\eref{Involutive} implies that
\begin{equation}\label{DefFrob}
[\text{im}\,A,\text{im}\,A]\subseteq\text{im}\,A.
\end{equation}
By definition, operator~\eref{WideOperator} is the variational anchor in the variational Lie algebroid~\cite{Galli10}; its total space is the quotient of the sum over~$J^\infty(\pi)$ of the initial domains for~$A_i$ by the kernel $\ker A=\bigl\{\boldsymbol{p}\in\Gamma\Omega\bigl(\xi_\pi\bigr)\mid \sum\nolimits_{i=1}^N A_i(\boldsymbol{p_i})=0\bigr\}$.
We emphasize that variational Lie algebroids are not Lie algebroids and the variational anchors are not anchors as
described 
by their traditional definition\footnote{The problem of integration of Lie algebroids to Lie groupoids, which is solved for manifolds (see~\cite{CrainicFernandes}), 
is open in the variational case over the jet bundles.}
\cite{Pradines1967,Vaintrob} 
for usual manifolds (here, for the target $M^m$~alone but not $J^\infty(\varSigma^n\to M^m)$). 

The operator~$A$ transfers the commutator~$[\,,\,]$ in the Lie algebra 
of evolutionary vector fields to the Lie algebra structure~$[\,,\,]_A$ on the quotient of its domain by the kernel~\cite{FulpLadaStasheffSH,Galli10}. It is now entirely obvious that for a given collection~$A_1$,\ $\ldots$,\ $A_N$ of the operators with a \emph{common} domain 
(and not necessarily involutive images, but this does not matter) the induced bracket~$[\,,\,]_A$ may not determine a well defined Lie structure on a single copy of that common domain (c.f.\ \cite{BarnichBialowieza}): Indeed, it is the concatenations~$\boldsymbol{p}$ of the $N$~elements~$\boldsymbol{p}_i$ from the same space for which the new bracket actually appears.

In fact, it is not the horizontal 
modules~$\Gamma\bigl(\pi_\infty^*(\xi_i)\bigr)$ 
of sections~$\boldsymbol{p}_i\bigl(\boldsymbol{x},
[\boldsymbol{q}]\bigr)$ 
of the induced fibre bundles $\pi_\infty^*(\xi_i)$ which we really need for the representation of the geometry at hand in terms of the homological vector fields, 
but it is the 
product $\overline{J^\infty}(\xi_\pi)\mathrel{{:}{=}}
J^\infty(\xi)\mathbin{{\times}_{\varSigma^n}}J^\infty(\pi)$
over~$\varSigma^n$ of the infinite jet spaces for~$\pi$
and~$\xi=\bigoplus\nolimits_{i=1}^N \xi_i$.
This means that we operate with the jet coordinates 
$\boldsymbol{p}_{i;\tau}$ corresponding to the multi\/-\/indices~$\tau$ instead of the derivatives $\tfrac{\mathrm{d}^{|\tau|}}{\mathrm{d}\boldsymbol{x}^\tau}\boldsymbol{p}_i\bigl(\boldsymbol{x},[\boldsymbol{q}]\bigr)$, whence the linear differential operators~$A_i$ become linear vector\/-\/functions of the jet variables~$\boldsymbol{p}_{i;\tau}$.
(In practice, this often means also that the number of the
``fields'' \emph{doubles} at exactly this moment.)
The following diagram endows the total space 
$\overline{J^\infty}(\xi_\pi)$ 
with the Cartan connection~$\nabla_{\mathcal{C}}$
(c.f.~\cite{Quillen1985}),
\begin{equation}\label{DiagConnection}
\begin{diagram}
J^\infty(\xi)\mathbin{{\times}_{\varSigma^n}}J^\infty(\pi) &
 \pile{\rTeXto^{\phantom{\pi_\infty^*(\xi_\infty)}} 
   \\    \lMiss_{
  \nabla_{\mathcal{C}}^\pi   
  } } 
  &
J^\infty(\pi) \\
\dTeXto \uMiss_{\nabla_{\mathcal{C}}^\xi} & & 
 \dTeXto^{\pi_\infty} \uMiss_{\nabla_{\mathcal{C}}^\pi} 
  \\
J^\infty(\xi) &
 \pile{\rTeXto^{\xi_\infty} \\ 
  \lMiss_{\nabla_{\mathcal{C}}^\xi} } 
  & \varSigma^n.
\end{diagram}
\end{equation}
This justifies the application of the total derivatives to the variables~$\boldsymbol{p}$ (including each~$\boldsymbol{p}_i$ alone) and, from now on, allows us to consider on 
$\overline{J^\infty}(\xi_\pi)$ the evolutionary vector fields with their sections depending (non)linearly on the fields~$\boldsymbol{q}$ and the variables~$\boldsymbol{p}_i$.

The 
conversion of the sections~$\boldsymbol{p}_i$ to jet variables creates the miracle: the kernel~$\ker A$ of the operator~$A$ 
becomes a linear subspace in~$\overline{J^\infty}(\xi_\pi)$ 
for each point of~$J^\infty(\pi)$. This is a significant achievement because it allows us to operate with the quotient spaces, which we anticipated when the bi\/-\/differential structure constants~$\mathbf{c}_{ij}^k$ were introduced in~\eref{BiDiffC}.

Next, we take the fibres of the vector bundles~$\xi_i$ (hence, of~$\xi_{i,\infty}$) and reverse their parity, $\Pi\colon\boldsymbol{p}_i\rightleftarrows\boldsymbol{b}_i$, the entire underlying jet space~$J^\infty(\pi)$ remaining intact.
This produces the horizontal infinite jet superbundle
$\overline{J^\infty}(\Pi\xi_\pi)\mathrel{{:}{=}}J^\infty(\Pi\xi)\mathbin{{\times}_{\varSigma^n}}J^\infty(\pi)\to J^\infty(\pi)$ with odd fibres, the coordinates there being~$\boldsymbol{b}_{i;\tau}$ for $i=1,\ldots,N$ and~$|\tau|\geq0$.
(In a special class of geometries, see section~\ref{SecGauge}, we shall recognize the variables~$\boldsymbol{b}_i$ as the \emph{ghosts}, which are denoted usually 
by~$\boldsymbol{\gamma}_i$.)
The operators~$A_i$ tautologically extend to~$\overline{J^\infty}(\Pi\xi_\pi)$ and become fibrewise\/-\/linear functions in~$\boldsymbol{b}_{i;\tau}$; the equivalence classes of the bi\/-\/differential Christoffel symbols~$\Gamma_{ij}^k$ are represented by 
bi\/-\/linear functions on that superspace.

\begin{theor}
Whatever representatives in the equivalence classes of the bi\/-\/differential symbols~$\Gamma_{ij}^k$ be taken, 
the odd\/-\/parity evolutionary vector field
\begin{equation}\label{Q}
\boldsymbol{Q}=\partial^{(\boldsymbol{q})}_{\sum\nolimits_{i=1}^N A_i(\boldsymbol{b}_i)} 
 -\frac{1}{2}\sum\limits_{k=1}^N 
  \partial^{(\boldsymbol{b}_k)}_{\sum\nolimits_{i,j=1}^N
 \Gamma_{ij}^k(\boldsymbol{b}_i,\boldsymbol{b}_j)}
\end{equation}
is homological\textup{:}
\[
\boldsymbol{Q}^2=0\pmod{\sum\limits_{k=1}^N
 \partial^{(\boldsymbol{b}_k)}_{\varphi_k(
   \boldsymbol{b}\otimes\boldsymbol{b}\otimes\boldsymbol{b})}
 \mid \sum\limits_{k=1}^N A_k(\varphi_k)=0}.
\]
\end{theor}

\noindent%
The proof is obtained immediately by passing from many arguments of many operators to the tall sections of wide operators, see~\eref{BothTallWide},
and then literally repeating the first half of the proof of the main theorem in~\cite{Galli10}.
However, we still prefer to prove the theorem straightforwardly; this will attest that the nature of the variables~$\boldsymbol{b}_i$ at different $i$'s may be entirely uncorrelated.

\begin{proof}
The halved anticommutator $\tfrac{1}{2}[\boldsymbol{Q},\boldsymbol{Q}]=\boldsymbol{Q}^2$ of the odd vector field~$\boldsymbol{Q}$ with itself is an evolutionary vector field.
In this anticommutator, the coefficient of~$\partial/\partial\boldsymbol{q}$, which 
determines the 
coefficients of~$\partial/\partial\boldsymbol{q}_\sigma$ at all~$|\sigma|\geq0$, is equal to
\begin{align*}
{}\partial^{(\boldsymbol{q})}_{\sum\nolimits_{i=1}^N
 A_i(\boldsymbol{b}_i)}& \Bigl(
\sum\limits_{j=1}^N  A_j(\boldsymbol{b}_j)\Bigr)
-\frac{1}{2}\sum\limits_{k,\ell=1}^N \delta_k^\ell\cdot
 A_\ell\Bigl(\sum\limits_{i,j=1}^N 
 \Gamma_{ij}^k(\boldsymbol{b}_i,\boldsymbol{b}_j)\Bigr);\\
\intertext{we now double the minuend --taking into account that $\boldsymbol{b}_i$ and~$\boldsymbol{b}_j$ are odd, whence the minus sign occurs after their 
interchange--}
{}&=\frac{1}{2}\sum\limits_{i,j=1}^N\Bigl(
\partial^{(\boldsymbol{q})}_{A_i(\boldsymbol{b}_i)}(A_j)(\boldsymbol{b}_j) - \partial^{(\boldsymbol{q})}_{A_j(\boldsymbol{b}_j)}(A_i)(\boldsymbol{b}_i)
-\sum\limits_{k=1}^N A_k\bigl(\Gamma^k_{ij}(\boldsymbol{b}_i,\boldsymbol{b}_j)\bigr)\Bigr)=0,
\end{align*}
by the definition of the symbols~$\Gamma_{ij}^k$.

Second, let us consider the Jacobi identity for the Lie algebra of evolutionary vector fields with the generating sections belonging to the images of the operators~$A_1$,\ $\ldots$,~$A_N$ viewed as the fibrewise\/-\/linear functions
on the total space of the bundle
$\overline{J^\infty}(\xi_\pi)\to J^\infty(\pi)$:
\begin{align}
0&=\sum\limits_{\substack{\circlearrowright \\ (imn)}}
\bigl[A_i(\boldsymbol{p}_i),
\sum\nolimits_k A_k\bigl(\Gamma_{mn}^k(\boldsymbol{p}_m,
\boldsymbol{p}_n)\bigr)\bigr]\notag \\
{}&\quad{}=\sum\limits_{\substack{\circlearrowright \\ (imn)}}
\sum\limits_k A_k\Bigl\{
\partial^{(\boldsymbol{q})}_{A_i(\boldsymbol{p}_i)}
 \bigl(\Gamma_{mn}^k(\boldsymbol{p}_m,\boldsymbol{p}_n)\bigr)
+\sum\limits_\ell \Gamma^k_{i\ell}\bigl(\boldsymbol{p}_i,
 \Gamma^\ell_{mn}(\boldsymbol{p}_m,\boldsymbol{p}_n)\bigr)
 \Bigr\},\notag\\ 
\intertext{where we have relabelled the indices $\ell\rightleftarrows k$ in the second sum,}
{}&\quad{}=-\sum\limits_{\substack{\circlearrowright \\ (imn)}}
\sum\limits_k A_k\Bigl\{
-\partial^{(\boldsymbol{q})}_{A_i(\boldsymbol{p}_i)}
 \bigl(\Gamma_{mn}^k(\boldsymbol{p}_m,\boldsymbol{p}_n)\bigr)
+\sum\limits_\ell \Gamma^k_{\ell i}\bigl(
 \Gamma^\ell_{mn}(\boldsymbol{p}_m,\boldsymbol{p}_n),
 \boldsymbol{p}_i\bigr)\Bigr\}.\label{JacobiGamma}
\end{align}
Thirdly, the velocity of each odd variable~$\boldsymbol{b}_k$ induced by the (for convenience, \emph{not} halved) anticommutator $[\boldsymbol{Q},\boldsymbol{Q}]=2\boldsymbol{Q}^2$ is obtained as follows,
\begin{align*}
-&\partial^{(\boldsymbol{q})}_{\sum\nolimits_{i=1}^N 
 A_i(\boldsymbol{b}_i)}
\bigl(\sum\limits_{m,n} \Gamma^k_{mn}(\boldsymbol{b}_m,\boldsymbol{b}_n)\bigr)
 +\frac{1}{2}\sum\limits_{\ell=1}^N
\partial^{(\boldsymbol{b}_\ell)}_{\sum\limits_{m,n}
 \Gamma^\ell_{mn}(\boldsymbol{b}_m,\boldsymbol{b}_n)}
\Bigl(\sum\limits_{j,i} \Gamma^k_{ji}(\boldsymbol{b}_j,\boldsymbol{b}_j)\Bigr) \\
{}&\quad{}= -\partial^{(\boldsymbol{q})}_{\sum\nolimits_{i=1}^N 
 A_i(\boldsymbol{b}_i)}
\bigl(\sum\limits_{m,n} \Gamma^k_{mn}(\boldsymbol{b}_m,\boldsymbol{b}_n)\bigr)
+\frac{1}{2}\sum\limits_{\ell,i} \Gamma^k_{\ell i}\Bigl(
\sum\limits_{m,n} \Gamma^\ell_{mn}(\boldsymbol{b}_m,\boldsymbol{b}_n),\boldsymbol{b}_i\Bigr)
-\frac{1}{2}\sum\limits_{i,\ell} \Gamma^k_{i\ell}\Bigl(
\boldsymbol{b}_i,\sum\limits_{m,n} 
 \Gamma^\ell_{mn}(\boldsymbol{b}_m,\boldsymbol{b}_n)\Bigr) \\
{}&\quad{}=\sum_{i,m,n}\Bigl\{ 
 -\partial^{(\boldsymbol{q})}_{A_i(\boldsymbol{b}_i)}
 \bigl(\Gamma^k_{mn}(\boldsymbol{b}_m,\boldsymbol{b}_n)\bigr)
+\sum\limits_{\ell=1}^N \Gamma^k_{\ell i}\bigl(
 \Gamma^\ell_{mn}(\boldsymbol{b}_m,\boldsymbol{b}_n),
   \boldsymbol{b}_i\bigr)\Bigr\}.\\
\intertext{We notice that the extra sum over the three cyclic permutations of each fixed set of the indices $i$,\ $m$, and~$n$ does not produce any change of the signs because the cyclic permutations (of the respective three odd $\boldsymbol{b}$'s) are even. Consequently, by taking the sum of all possible Jacobi identities~\eref{JacobiGamma}, we conclude that the sought\/-\/for coefficient is equal to}
{}&\quad{}=\frac{1}{3} \sum\limits_{i,m,n}
 \sum\limits_{\substack{\circlearrowright \\ (imn)}}
\Bigl\{\dots\Bigr\} = 0\pmod{
\boldsymbol{\varphi}=(\varphi_1,\ldots,\varphi_N) \mid
\sum\limits_{k=1}^N A_k(\varphi_k)=0}.
\end{align*}
This completes the proof.
\end{proof}

We remark that the above construction of the homological vector fields~$\boldsymbol{Q}$ in absence of the equations of motion but with an \textit{a priori} given collection~$A_1$,\ $\ldots$,\ $A_N$ of generators~\eref{Q} for the infinitesimal transformations of the model patterns upon the construction of nontrivial gauge theories with the zero action functional and the prescribed nontrivial gauge group, 
see~\cite{BaulieuSinger1988,Schwarz1979}.


\section{Examples: Three regular classes}\label{SecExamples}
\subsection{Variational Poisson algebroids}\label{SecHam}
The definition of Hamiltonian operators on jet spaces~\cite{GelfandDorfman}, which are a class of linear differential operators in total derivatives, reads as follows~\cite{Olver,Lstar}.
Consider a total differential operator $A\colon\widehat{\varkappa(\pi)}\to\varkappa(\pi)$ that maps variational covectors to evolutionary fields. Let $\mathcal{H}_1$,\ $\mathcal{H}_2\in\overline{H}^n(\pi)$ be two Hamiltonian functionals, that is, the equivalence classes of two elements in the coupling module~$\overline{\Lambda}^n(\pi)$ of the highest horizontal forms. The operator~$A$ is \emph{Hamiltonian} if the bracket~$\{\,,\,\}_A$ defined by the formula
\[
\{\mathcal{H}_1,\mathcal{H}_2\}_A\mathrel{{:}{=}}
\left<\frac{\delta\mathcal{H}_1}{\delta\boldsymbol{q}},
A\Bigl(\frac{\delta\mathcal{H}_2}{\delta\boldsymbol{q}}\Bigr)\right>,\qquad
\langle\,,\,\rangle\colon\widehat{\varkappa(\pi)}\times
\varkappa(\pi)\to\overline{\Lambda}^n(\pi),
\]
is a Poisson structure, i.e., if it is bi\/-\/linear (which comes automatically), is skew\/-\/symmetric (whence $A=-A^\dagger$; the $\mathbb{Z}_2$-\/graded case is more subtle, c.f.~\cite{Criterion}), and satisfies the Jacobi identity.
There are many formulations of convenient criteria for the verification of the latter (see~\cite{Olver,GelfandDorfman} and also~\cite{d3Bous,Criterion}). It is important that they reveal --or are based effectively on-- the commutation closure~\eref{DefFrob} for the images of Hamiltonian operators.

For a Hamiltonian operator thus defined on the empty jet space~$J^\infty(\pi)$, a determined autonomous evolution equation is called \emph{Hamiltonian} with respect to that given structure if the equation can be cast into the form
\[
\mathcal{E}=\bigl\{\boldsymbol{F}=\boldsymbol{q}_t-A\left(
 \frac{\delta\mathcal{H}}{\delta\boldsymbol{q}}\right)=0\mid
\mathcal{H}\in\overline{H}^n(\pi)\bigr\}.
\]
The left\/-\/hand side(s) of the equation~$\mathcal{E}$ belong to a suitable horizontal module~$P_0$ of form~$\Gamma\bigl(\pi_\infty^*(\xi)\bigr)$ for some~$\xi$;
\emph{usually}, this module is identified with the module 
$\varkappa(\pi)=\Gamma\bigl(\pi_\infty^*(\pi)\bigr)$
of the velocities.
This identification 
conveniently fills in several small gaps in the count of the base dimensions,
the location of the Hamiltonian~$\mathcal{H}$ for~$\mathcal{E}$ in the highest horizontal cohomology group but not in the preceding one, and 
the (in/ex)clu\-si\-on of the time\/-\/derivatives of~$\boldsymbol{q}$ in the module~$P_0$ of equations. However, we emphasize that the reparametrizations of the unknowns~$\boldsymbol{q}$ and equations~$\boldsymbol{F}=0$ are \emph{not} correlated in principle. Let us remember that the arguments~$\boldsymbol{p}$ of the Hamiltonian operator for~$\mathcal{E}$ belong to the module~$\widehat{P_0}$ which is $\langle\,,\,\rangle$-\/dual to~$P_0$. While the system~$\mathcal{E}$ remains in evolutionary transcription, the variational derivatives $\delta\varrho/\delta\boldsymbol{q}$ of conserved densities~$\varrho$ for~$\mathcal{E}$ constitute a linear subspace of~$\widehat{P_0}$, but this correspondence vanishes under more general reparametrizations of the equations.

Reversing the parity of the sections~$\boldsymbol{p}$, we arrive at the following diagram:
\[ 
\begin{diagram}
\text{antifields} & \rTeXto & \boldsymbol{b}\in\Pi\widehat{P_0} & & \\
{} & & \uTeXto^{\Pi} \dTeXto & & {}\\
{} & & \boldsymbol{p}\in\widehat{P_0} & 
 \rBothWays^{*}_{\phantom{MM}\langle\,,\,\rangle\phantom{MM}} & 
 \boldsymbol{F}\in P_0\simeq\varkappa(\pi)\longleftarrow
\lefteqn{\text{usual identification.}}
\end{diagram}
\]
We remark that, under the labelling of the evolution equations $\mathcal{E}=\{\boldsymbol{q}_t=\text{r.-h.s.}\}$
by the fields~$\boldsymbol{q}$ themselves, the \emph{antifields} are then, on these grounds, usually denoted\footnote{Without and with regard to the underlying metric and the Hodge structure, respectively (c.f.~\cite{CattaneoFelder2000} for an example); because we do not have any metric involved explicitly in the model, we entirely discard this distinction.}
by~$\boldsymbol{q}^\dagger$ or~$\boldsymbol{q}^*$. But in view of the non\/-\/identical correspondence between the fields~$\boldsymbol{q}$ and the left\/-\/hand sides~$\boldsymbol{F}$ of the equations $\mathcal{E}=
\{\boldsymbol{F}=0\}$, 
the proper notation for the antifields would 
be~$\boldsymbol{F}^\dagger$ (resp.,~$\boldsymbol{F}^*$).

Not only the images of Hamiltonian operators are involutive so that
\[
\bigl[A(\boldsymbol{p}_1),A(\boldsymbol{p}_2)\bigr]=
A\bigl([\boldsymbol{p}_1,\boldsymbol{p}_2]_A\bigr),
\]
but the brackets~$[\,,\,]_A$ are calculated effectively for each given~$A$, see, e.g.,~\cite{d3Bous,Criterion}. Therefore, we have all the data of section~\ref{SecInvolutive}
ready at hand: 
taking the odd jet variables~$\boldsymbol{b}=\Pi(\boldsymbol{p})$, 
we obtain the odd evolutionary vector fields
\begin{equation}\label{QPoisson}
\boldsymbol{Q}=\partial^{(\boldsymbol{q})}_{A(\boldsymbol{b})}
-\frac{1}{2}\partial^{(\boldsymbol{b})}_{\Gamma^A_{AA}(\boldsymbol{b},\boldsymbol{b})},
\end{equation}
which encode the variational Poisson structures (primarily, the Jacobi identity, see above) in terms of the homological condition~$\boldsymbol{Q}^2=0$.

The homological vector fields~$\boldsymbol{Q}$ for Hamiltonian operators were constructed in the recent paper~\cite{JKGolovko2008} by taking the advantage of the known geometric interpretation for~$\boldsymbol{b}$ as the parity\/-\/reversed variational covectors --but without 
references to variational Poisson algebroids. Namely, this was done by explicitly calculating the induced velocities of the sections~$\boldsymbol{p}\bigl(\boldsymbol{x},[\boldsymbol{q}]\bigr)$ for \textit{a priori} specified velocities~$\varphi=A(\boldsymbol{p})$ of the unknowns~$\boldsymbol{q}$; we shall apply the same technique in section~\ref{SecGauge}.

\begin{example}
There are many examples when the Hamiltonian operators~$A$
are addressed in the context of the brackets~$[\,,\,]_A$ and
bi\/-\/differential symbols~$\Gamma^A_{AA}$. For instance,
the notorious KdV equation (say, upon $w(x,t)$) yields
(\cite{SokolovUMN}, see also the Appendix)
\begin{equation}\label{QKdV}
A_2^{\text{KdV}}=-\tfrac{1}{2}\tfrac{\mathrm{d}^3}{\mathrm{d}x^3}+2w\tfrac{\mathrm{d}}{\mathrm{d}x}+w_x,\qquad
\Gamma^{\cdot}_{\cdot,\cdot}(p^1,p^2)=\tfrac{\mathrm{d}}{\mathrm{d}x}(p^1)\cdot p^2-p^1\cdot\tfrac{\mathrm{d}}{\mathrm{d}x}(p^2),\qquad
\boldsymbol{Q}=\partial^{(w)}_{A_2^{\text{KdV}}(b)}
 +\partial^{(b)}_{bb_x},
\end{equation}
which is the minimally possible nontrivial illustration;
we refer to~\cite{TMPhGallipoli} for the Boussinesq system,
the Kaup\/--\/Boussinesq equation was considered in~\cite{KarabanovaKiselev}; the Hamiltonian operators and the bi\/-\/differential Christoffel symbols for the Drinfeld\/--\/Sokolov equations associated with the root systems of rank two are contained in~\cite{Protaras2008}.
\end{example}

The calculation of variational Poisson cohomology groups determined by the evolutionary 
Poisson differentials~$\boldsymbol{Q}$ is a rapidly developing topic, see~\cite{Getzler,
KacJune2011}. We regret 
to notice the extent to which the progress in the cohomology theory for variational Poisson algebroids is retarded --by three decades at the least-- in comparison with the BRST-{} and BV\/-\/cohomology technique for the gauge algebroids (see section~\ref{SecGauge}).

The evolutionary vector field~$\boldsymbol{Q}$ in~\eref{QPoisson} is itself Hamiltonian with respect to the even $W$-\/charge $\bar{\boldsymbol{\Omega}}=-\tfrac{1}{2}\langle\boldsymbol{b},A(\boldsymbol{b})\rangle$ and the canonical symplectic structure $\omega=\left(\begin{smallmatrix}\phantom{+}0 & 1\\
-1 & 0\end{smallmatrix}\right)$ 
that stems from the volume form~$\delta\boldsymbol{p}\wedge\delta\boldsymbol{q}$ in the fibres over points of~$\varSigma^n$ in the variational cotangent bundle~\cite{KuperCotangent} to the jet space~$J^\infty(\pi)$:
\[
\boldsymbol{Q}=\partial^{(\boldsymbol{q})}_{\textstyle\delta\bar{\boldsymbol{\Omega}}/\delta\boldsymbol{b}} +
\partial^{(\boldsymbol{b})}_{\textstyle -\delta\bar{\boldsymbol{\Omega}}/\delta\boldsymbol{q}}.
\]
We emphasize that this property of the field~$\boldsymbol{Q}$ associated with a Hamiltonian operator~$A$ does not require the target~$M$ in the initial geometric setup $J^\infty(\varSigma^n\to M^m)$ to be a Poisson manifold, c.f.~\cite{AKZS,CattaneoFelder2000,FulpLadaStasheffSrni}.

The condition $\boldsymbol{Q}^2=\tfrac{1}{2}[\boldsymbol{Q},
\boldsymbol{Q}]=0$ upon the evolutionary vector field~$\boldsymbol{Q}$ is equivalent to the variational classical master equation $[\![\boldsymbol{\Omega},\boldsymbol{\Omega}]\!]=0$ upon the variational Poisson bi\/-\/vector 
$\boldsymbol{\Omega}=\tfrac{1}{2}\langle\boldsymbol{b},A(\boldsymbol{b})\rangle$ and the variational Schouten bracket 
$[\![\boldsymbol{\omega}_1,\boldsymbol{\omega}_2]\!]=
\left<\overrightarrow{\delta}\boldsymbol{\omega}_1\wedge
\overleftarrow{\delta}\boldsymbol{\omega}_2\right>$
(the odd\/-\/parity Poisson bracket or the
\emph{antibracket}, see~\cite{ZinnJustin1975} or~\cite{Memorandum1641} and references therein), which in the adopted notation and with Dirac's convention reads
\begin{align*}
[\![\boldsymbol{\omega}_1,\boldsymbol{\omega}_2]\!]&=
\Bigl\langle \Bigl(\frac{\overrightarrow{\delta}\boldsymbol{\omega}_1}{\delta\boldsymbol{q}} \delta\boldsymbol{q} +
\frac{\overrightarrow{\delta}\boldsymbol{\omega}_1}{\delta\boldsymbol{b}} \delta\boldsymbol{b}\Bigr) \wedge
\Bigl(\delta\boldsymbol{q} \frac{\overleftarrow{\delta}\boldsymbol{\omega}_2}{\delta\boldsymbol{q}} +
\delta\boldsymbol{b} \frac{\overleftarrow{\delta}\boldsymbol{\omega}_2}{\delta\boldsymbol{b}} \Bigr)
\Bigr\rangle \\
{}&= \frac{\overrightarrow{\delta}\boldsymbol{\omega}_1}{\delta\boldsymbol{b}} \langle\delta\boldsymbol{b},
\delta\boldsymbol{q}\rangle \frac{\overleftarrow{\delta}\boldsymbol{\omega}_2}{\delta\boldsymbol{q}}
- \frac{\overrightarrow{\delta}\boldsymbol{\omega}_1}{\delta\boldsymbol{q}} \langle\delta\boldsymbol{b},
\delta\boldsymbol{q}\rangle \frac{\overleftarrow{\delta}\boldsymbol{\omega}_2}{\delta\boldsymbol{b}}
=
\left[
\frac{\overrightarrow{\delta}\boldsymbol{\omega}_1}{\delta\boldsymbol{b}}\cdot
\frac{\overleftarrow{\delta}\boldsymbol{\omega}_2}{\delta\boldsymbol{q}} -
\frac{\overrightarrow{\delta}\boldsymbol{\omega}_1}{\delta\boldsymbol{q}}\cdot
\frac{\overleftarrow{\delta}\boldsymbol{\omega}_2}{\delta\boldsymbol{b}}\right],
\end{align*}
where the brackets~$[\ ]$ mark the cohomology class \emph{after} the volume form~$\mathrm{d}\text{vol}\,\varSigma^n$ is fixed in the coupling~$\langle\,,\,\rangle$.
The composition of the 
antibracket\footnote{Let us recall that the 
Schouten bracket of variational one\/-\/vectors $\langle\boldsymbol{b},\varphi_i\rangle$ calculates the commutator
$[\varphi_1,\varphi_2]=\partial^{(\boldsymbol{q})}_{\varphi_1}(\varphi_2)-\partial^{(\boldsymbol{q})}_{\varphi_2}(\varphi_1)$ of the two evolutionary vector fields~$\partial^{(\boldsymbol{q})}_{\varphi_i}$: Namely, we have that
$[\![\langle\boldsymbol{b},\varphi_1\rangle,
\langle\boldsymbol{b},\varphi_2\rangle]\!]=
\langle\boldsymbol{b},[\varphi_1,\varphi_2]\rangle$.
This formula illustrates our earlier warning about the absence of any reasonable Leibniz rule for either~$[\,,\,]$ or~$[\![\,,\,]\!]$.}
relies on the arrangement of the jet 
(super)\/fibre variables in the coupling\/-\/dual pairs of opposite parity: here, we have the fields~$\boldsymbol{q}$ 
and the antifields~$\boldsymbol{b}=\boldsymbol{q}^\dagger$ so that (up to the minus sign in~$\bar{\boldsymbol{\Omega}}
=-\boldsymbol{\Omega}$ which is due to Dirac's convention)
\[
\boldsymbol{Q}=\partial^{(\boldsymbol{q})}_{\textstyle[\![\bar{\boldsymbol{\Omega}},\boldsymbol{q}]\!]} +
\partial^{(\boldsymbol{q}^\dagger)}_{\textstyle[\![
\bar{\boldsymbol{\Omega}},\boldsymbol{q}^\dagger]\!]}
\qquad \text{so that}\quad
\boldsymbol{Q}(\Xi)=[\![\bar{\boldsymbol{\Omega}},\Xi]\!]\quad
\forall\ \Xi.
\]
Gauge models, which we address in section~\ref{SecGauge},
provide a different geometric interpretation for the antifields and require the introduction of at least one generation of the ghost\/-\/antighost pairs.

The definition of variational Poisson structures admits its further generalization to \emph{non}\/-\/evolutionary systems~$\mathcal{E}$.
While the bi\/-\/linearity of the bracket remains intact, 
its skew\/-\/symmetry imposes a nontrivial constraint upon the operator~$A\colon\widehat{P_0}\to\text{sym}\,\mathcal{E}$ in terms of the linearization of the system~$\mathcal{E}$ at hand, whereas the Jacobi identity amounts to the same condition~$\boldsymbol{Q}^2=0$ upon the odd vector fields~\eref{QPoisson}. A significant progress in this direction has been achieved only very recently, see~\cite{JK2011WDVV,Topical}. 

\begin{prb}
Do there exist Hamiltonian operators~$A_1$,\ $\ldots$,\ $A_N$ which are compatible in the sense of~\eref{Involutive}, and what then could be the geometry standing behind the bi\/-\/differential Christoffel symbols (c.f.~\cite{Christoffel})\,?
\end{prb}

Finally, we recall that the root systems of semi\/-\/simple complex Lie algebras are a regular source of variational Poisson structures for KdV\/-\/type models~\cite{DSViniti84}; this is achieved by using the 2D~Toda chains.

\subsection{2D~Toda chains}\label{SecToda}
The hyperbolic two\/-\/dimensional non\/-\/periodic Toda chains~\cite{LeznovSaveliev1979,YamilovShabat} associated with the root systems of semi\/-\/simple complex Lie algebras~$\mathfrak{g}$ of rank~$r$ are the exponential\/-\/nonlinear Euler\/--\/Lagrange systems
\[
q^i_{xy}=\exp\bigl(K^i_{\,j}q^j\bigr),\qquad 1\leq i\leq r,
\]
where $K$~is the Cartan matrix of~$\mathfrak{g}$.
The action functionals for the 2D~Toda chains are expressed in terms of the roots and are non\/-\/polynomial, thus exceeding the assumptions of~\cite{FulpLadaStasheffSH}.
Also, it is very instructive to inspect the non\/-\/identical correlation between the Noether symmetries and the generating functions of conservation laws for these Lagrangian equations, see~\cite{TMPhGallipoli,SymToda} for detail.

The 2D~Toda chains are the best representatives of the vast class of Euler\/--\/Lagrange systems of Liouville type 
(see~\cite{SokolovUMN,SokStar} and references therein, also~\cite{TMPhGallipoli,SymToda}). By definition, the systems of this type possess as many Liouville's \emph{integrals} $w\bigl(x,y;[\boldsymbol{q}]\bigr)$ such that $\tfrac{\mathrm{d}}{\mathrm{d}y}(w)\approx0$ on\/-\/shell and similarly, $\overline{w}\bigl(x,y;[\boldsymbol{q}]\bigr)$ satisfying
$\tfrac{\mathrm{d}}{\mathrm{d}x}(\overline{w})\approx0$, as there are unknowns~$\boldsymbol{q}$. For example, consider the Liouville equation $\mathcal{E}=\bigl\{F\equiv q_{xy}-\exp(2q)=0\bigr\}$ and let $w=q_x^2-q_{xx}$ so 
that~$\tfrac{\mathrm{d}}{\mathrm{d}y}(w)=-\tfrac{\mathrm{d}}{\mathrm{d}x}(F)$.

We have shown in~\cite{TMPhGallipoli} that the integrals for this class of Euler\/--\/Lagrange equations in fact depend differentially on the momenta~$\mathfrak{m}$ which emerge from the kinetic part of the action; this allows us to cast~$\mathcal{E}$ into the evolutionary form $q_y=\delta\mathrm{H}/\delta\mathfrak{m}$, $\mathfrak{m}_y=-\delta\mathrm{H}/\delta q$. The differential orders of the integrals $w=w[\mathfrak{m}]$ the concide with the exponents of the Lie algebra~$\mathfrak{g}$, see~\cite{YamilovShabat,SymToda}; there are constructive procedures to obtain the integrals~$w$ and~$\overline{w}$ (see~\cite{Saveliev1992}
and~\cite{Shabat95} illustrated in~\cite{Protaras2008}).

In~\cite{TMPhGallipoli,SymToda} we proved that the adjoint linearizations of the integrals with respect to the momenta,
\begin{equation}\label{Square}
\square={\bigl(\ell^{(\mathfrak{m})}_w\bigr)}^\dagger
\end{equation}
are linear differential operators with involutive images, see~\eref{DefFrob}. These operators produce symmetries of the Liouville\/-\/type system at hand (in particular, its Noether symmetries $\varphi_{\mathcal{L}}=\delta\mathcal{H}\bigl[w[\mathfrak{m}]\bigr]/\delta\mathfrak{m}=\square\bigl(\delta\mathcal{H}[w]/\delta w\bigr)$). 
By construction, each column of the operator~$\square$ stems from the respective integral for~$\mathcal{E}$.
However, the image of a particular column may not itself be closed under commutation: we encounter an example for the root system~$\mathsf{A}_2$ (here, for the second, higher\/-\/order integral and, respectively, the second column in~\eref{Square}, see~\cite{TMPhGallipoli,Protaras2008}). 

The geometric nature of the domains of the operators~$\square$ is the most nontrivial in comparison with the other two sample geometries which we discuss in sections~\ref{SecHam} and~\ref{SecGauge} of this review. In brief, the arguments~$\boldsymbol{p}\bigl(x,\bigl[w[\mathfrak{m}]\bigr]\bigr)$ in each such domain belong to the variational cotangent bundle to an equation \emph{other than}~$\mathcal{E}$, while the Miura 
substitutions~$w=w[\mathfrak{m}]$ relate the systems; we refer to~\cite{TMPhGallipoli} and~\cite{SymToda} for detail.
This reasoning implies that the components of the sections~$\boldsymbol{p}$ do not exist individually as well\/-\/defined ``functions'' and therefore the operators~$\square$ may not be split in separate columns. (Let us remember that the operators whose domains appear from the variational 
\emph{co}\-tan\-gent bundles~\cite{KuperCotangent}, like the Hamiltonian operators, were referred to in~\cite{SymToda,Galli10} as the operators of \emph{second~kind}.)

The term ``Miura'' is indeed justified in this context~\cite{DSViniti84}:    
the brackets~$[\,,\,]_\square$ on the domains of~$\square$ are calculated by using the correlation~\cite{SymToda} of the entire geometry of the 2D~Toda chains with the previous case of variational Poisson algebroids (section~\ref{SecHam}). The correspondence yields the symbols~$\Gamma^\square_{\square\square}$ explicitly in terms of the integrals~$w$ and the roots of~$\mathfrak{g}$; this calculates the commutation relations in the symmetry algebras for the 2D~Toda chains.

As usual, we reverse the parity of the arguments~$\boldsymbol{p}$ for the operators~$\square$ by taking $\Pi\colon\boldsymbol{p}\rightleftarrows\boldsymbol{b}$ and obtain the homological evolutionary vector field
\begin{equation}\label{QSquare}
\boldsymbol{Q}=\partial^{(\boldsymbol{q})}_{\square(\boldsymbol{b})}-\frac{1}{2}
\partial^{(\boldsymbol{b})}_{\Gamma^\square_{\square\square}(\boldsymbol{b},\boldsymbol{b})},\qquad \boldsymbol{Q}^2=0.
\end{equation}

\begin{example}  
The operator~$\square=q_x+\tfrac{1}{2}\tfrac{\mathrm{d}}{\mathrm{d}x}$ for the Liouville equation $q_{xy}=\exp(2q)$ specifies the differential~$\boldsymbol{Q}=\partial^{(q)}_{\square(b)}+\partial^{(b)}_{bb_x}$; the equality of the even velocity~$bb_x$ of the odd variable~$b$ to the respective velocity which we obtained in~\eref{QKdV} for the Korteweg\/--\/de Vries equation is no coincidence~\cite{SymToda}.
\end{example}

\begin{prb}
Is it true that for each operator~$\square$ of \emph{second kind} and for the associated differential~$\boldsymbol{Q}$ there is the \emph{Hamiltonian} operator~$A$ with the same domain such that~$\Gamma^\square_{\square\square}=\Gamma^A_{AA}$\,?
\end{prb}


\smallskip
Finally, we recall that the 2D~Toda chains related to the root systems are obtained by imposing the cylindric symmetry reduction in the Yang\/--\/Mills equations, the distinguished gauge models~\cite{LeznovSaveliev1979}. In turn, the latter do carry their own homological vector fields; we discuss them in the next section. However, the Yang\/--\/Mills equations are a class of models for which (unlike, e.g., gravity) the bi\/-\/differential Christoffel symbols standing in the fields~$\boldsymbol{Q}$ are structurally simple. This shows the ``preservation of complexity'' in its genuine sense of logic under the reduction (large,\ simple)~$\mapsto$ (small,\ complex) of the Yang\/--\/Mills systems to the 2D~Toda chains.
It is interesting to track the details of the correspondence between the homological vector fields~$\boldsymbol{Q}$,
the compositions of the Schouten bracket, and the respective
syzygies and charges in the two models and by this, to approach the origin of Liouville's integrals~$w$ along a new direction. This will give us a clue to their analogs in the affine case of~$\mathfrak{g}^{(k)}$.

\subsection{Gauge systems}\label{SecGauge}
The Euler\/--\/Lagrange systems $\mathcal{E}=\bigl\{
\boldsymbol{F}=\delta S/\delta\boldsymbol{q}=0\mid
S\in\overline{H}^n(\pi)\bigr\}$ contain as many equations as there are unknowns~$\boldsymbol{q}$. By construction, such equations $F_i=0$ are \emph{conveniently} labelled by the respective fields~$q^i$; let us remember that the horizontal module~$P_0$ of the sections~$F$ is then~$P_0\simeq\widehat{\varkappa(\pi)}$.
Because of this, the generating sections $\psi\in\widehat{P_0}$ of conservation laws for~$\mathcal{E}$ acquire the nature of Noether symmetries~$\varphi_{\mathcal{L}}\in\text{sym}\,\mathcal{E}$, which is indeed the case by virtue of the First Noether Theorem (up to, possibly, the non\/-\/identical Noether maps $\widehat{P_0}\to\varkappa(\pi)$, as we recalled in section~\ref{SecToda}).

Although the systems~$\mathcal{E}$ are determined, there may appear the differential constraints (also called \emph{syzygies}, \emph{Noether identities}, or \emph{Bianchi identities}) between the equations of motion,
\[
\Phi[\boldsymbol{F}]\equiv0\quad\forall\ \boldsymbol{q}=
\boldsymbol{q}(x),\qquad \Phi\in P_1.
\]
The relations $\Phi_{i+1}[\Phi_i]\equiv0$ between the relations, valid identically for all~$\Phi_{i-1}$ 
(here $\Phi_0=\boldsymbol{F}\in P_0$) give rise to possibly several but finitely many generations of the horizontal modules $P_i\ni\Phi_i$ for~$i>0$. In the sequel, we shall assume that the given action functional~$S$ determines the system~$\mathcal{E}=\{\boldsymbol{F}=0\}$ of equations of motion with one generation of the constraints~$\Phi[\boldsymbol{F}]=0$ and that there are no further relations between the already known ones. 
In the following diagram we summarize the notation\footnote{Again, we do not fix any metric on~$\varSigma^n$ and therefore 
not grasp the subtle difference between the transcripts~$\gamma^\dagger$ and~$\gamma^*$ for the antighosts (see, e.g.,~\cite{CattaneoFelder2000}).
We repeat that 
we do not pay any 
\emph{particular} attention to the upper or lower location of indices.}
and interpret 
the objects at hand in terms of the BV-\/theory~\cite{BV}:
\[
\begin{diagram}
\phantom{MMMMMl}\text{ghosts} & \rTeXto & 
  \boldsymbol{\gamma}_i=\boldsymbol{b}_i\in
 \Pi\widehat{P_1} & & 
  \boldsymbol{\gamma}_i^\dagger\in P_1 & \lTeXto &
\lefteqn{\text{antighosts}}\phantom{\text{usual identification.}} \\
{} & & \uTeXto^{\Pi} \dTeXto & & \uBothWays_{\text{id}} & & \\
\phantom{ll}
\text{gauge parameters} & \rTeXto & \boldsymbol{\epsilon}_i=
 \boldsymbol{p}_i\in\widehat{P_1} & \rBothWays^{*}_{\langle\,,\,\rangle} & \Phi^i[\boldsymbol{F}]\in P_1 & \lTeXto &
 \lefteqn{\text{Noether identities}}\phantom{\text{usual identification.}} \\
{} & & & & \boldsymbol{q}^\dagger\in\Pi P_0 & \lTeXto &
 \lefteqn{\text{antifields}}\phantom{\text{usual identification.}} \\
{} & & & & \uTeXto^{\Pi} \dTeXto & & {}\\
\text{Noether symmetries} & \rTeXto & \varphi_{\mathcal{L}}
\doteq\psi\in\varkappa(\pi)\simeq\widehat{P_0} & \rBothWays^{*}_{\phantom{M}\langle\,,\,\rangle\phantom{M}} & \boldsymbol{F}\in P_0\simeq\widehat{\varkappa(\pi)} & \lTeXto & \text{usual identification.}
\end{diagram}
\]
Each relation $\Phi^i[\boldsymbol{F}]=0$ between the equations~$F_j=0$, $1\leq j\leq m$, yields the linear total differential operator~$A_i\colon\widehat{P_1}\to\text{sym}\,\mathcal{E}\subset\varkappa(\pi)$ that generates symmetries of the system~$\mathcal{E}$. These symmetries are parametrized by arbitrary sections~$\boldsymbol{p}_i\bigl(\boldsymbol{x},[\boldsymbol{q}]\bigr)\in\widehat{P_1}$ and are called \emph{gauge symmetries}.

Namely, suppose that the identity $\Phi\bigl[\boldsymbol{F}[\boldsymbol{q}]\bigr]\equiv0$ holds irrespective of a section~$\boldsymbol{q}=\boldsymbol{q}(\boldsymbol{x})$. Consequently, this identity is indifferent to arbitrary infinitesimal shifts, which are given by the $\pi_\infty$-\/vertical evolutionary vector fields $\partial^{(\boldsymbol{q})}_\varphi$ on~$J^\infty(\pi)$. The chain rule implies that
\begin{align*}
\partial^{(\boldsymbol{F})}_{\textstyle \partial^{(\boldsymbol{q})}_\varphi (\boldsymbol{F})} (\Phi) &\equiv0\in P_1.\\
\intertext{Passing to the linearizations (the definition of which in the non\/-\/graded case is $\ell^{(\boldsymbol{a})}_\psi\bigl(\varphi[\boldsymbol{a}]\bigr)=\partial^{(\boldsymbol{a})}_\varphi(\psi)$ whenever this right\/-\/hand side is well\/-\/defined for the given~$\boldsymbol{a}$ and~$\psi$),
we conclude that}
\ell^{(\boldsymbol{F})}_\Phi \circ \ell^{(\boldsymbol{q})}_{\boldsymbol{F}} (\varphi) &=0,\\
\intertext{where the composition is a linear mapping from 
$\varkappa(\pi)$ to~$P_1$. Let us now couple this zero velocity along~$P_1$ with \emph{any} element~$\boldsymbol{p}\in\widehat{P_1}$ from the dual module and then integrate by parts, staying in the equivalence class of the zero in the cohomology. We obtain}
{\bigl(\ell^{(\boldsymbol{q})}_{\boldsymbol{F}}\bigr)}^\dagger
 \circ
{\bigl(\ell^{(\boldsymbol{F})}_{\Phi}\bigr)}^\dagger
 (\boldsymbol{p}) &=0.
\end{align*}
But let us recall that the linearization
\begin{align*}
\ell^{(\boldsymbol{q})}_{\boldsymbol{F}} &\colon
 \varkappa(\pi)\to P_0\simeq\widehat{\varkappa(\pi)}\\
\intertext{and the adjoint linearization}
{\bigl(\ell^{(\boldsymbol{q})}_{\boldsymbol{F}}\bigr)}^\dagger
 &\colon
 \widehat{P_0}\simeq\varkappa(\pi)\to 
  \widehat{\varkappa(\pi)}\simeq P_0
\end{align*}
\emph{coincide} for the Euler\/--\/Lagrange systems~$\mathcal{E}=\bigl\{\boldsymbol{F}=0\bigr\}$ due to the Helmholz criterion (see~\cite{Olver,Vin1984})
\[
\ell^{(\boldsymbol{q})}_{\boldsymbol{F}} =
{\bigl(\ell^{(\boldsymbol{q})}_{\boldsymbol{F}}\bigr)}^\dagger
\quad \Longleftrightarrow\quad
\exists\ S\in\overline{H}^n(\pi)\mid\boldsymbol{F}=\delta S/\delta\boldsymbol{q}.
\]
This implies that ${\bigl(\ell^{(\boldsymbol{F})}_{\Phi}\bigr)}^\dagger(\boldsymbol{p})$ is, on\/-\/shell, a symmetry of the system~$\mathcal{E}$ for any section~$\boldsymbol{p}\in\widehat{P_1}$. Moreover, for each Noether identity
$\Phi^i[\boldsymbol{F}]\in P_1$ we have constructed explicitly the operator
\[
A_i\mathrel{{:}{=}}{\bigl(\ell^{(\boldsymbol{F})}_{\Phi^i}\bigr)}^\dagger
\]
that yields the symmetries~$\varphi=A_i(\cdot)$ of the model.

We note that the constraints~$\Phi$ need not be linear.
Furthermore, in the cases of a very specific geometry of~$\mathcal{E}$ there may appear the symmetry\/-\/producing operators which 
do not issue 
from any differential relations between the equations of motion.
The approach which we have formulated by so far treats all such structures in a uniform~way.

However, under the additional assumption that the symmetries obtained in the images of the operators are Noether,
$\partial^{(\boldsymbol{q})}_{A(\boldsymbol{p})}(S)=[0]\in\overline{H}^n(\pi)$, the existence of the respective sections~$\Phi$ is justified easily and besides, 
these constraints appear to be \emph{linear}. Actually, 
suppose that $\partial^{(\boldsymbol{q})}_{A(\boldsymbol{p})}(S)=\bigl\langle\delta S/\delta\boldsymbol{q},A(\boldsymbol{p})\bigr\rangle=\bigl\langle\boldsymbol{p},A^\dagger(\boldsymbol{F})\bigr\rangle=[0]$ for all sections~$\boldsymbol{p}\in\widehat{P_1}$. This implies that the Noether identity $\Phi[\boldsymbol{F}]=0$ amounts to the linear differential relation~$A^\dagger(\boldsymbol{F})=0$ between the equations of motion so that $\Phi$~coincides with its own linearization.

\begin{example} 
The Maxwell equations $\mathcal{E}=\bigl\{F^i\equiv\partial_j\mathcal{F}^{ij}=0\bigr\}$ upon the skew\/-\/symmetric field strength tensor $\mathcal{F}_{ij}=\partial_i\mathcal{A}_j-\partial_j\mathcal{A}_i$ satisfy the obvious relation $\Phi=\partial_i F^i=0$ (that is, 
$\Phi=\bigl\{\partial_i\partial_j\mathcal{F}^{ij}\equiv0\bigr\}$), which manifests the invariance of the system (including its action functional) under (finite or infinitesimal) gauge transformations $\boldsymbol{\mathcal{A}}\mapsto\boldsymbol{\mathcal{A}}-\mathrm{d}_{\text{dR}}
\bigl(p(\boldsymbol{x},[\boldsymbol{\mathcal{A}}])\bigr)$ 
of the electromagnetic field~${\boldsymbol{\mathcal{A}}=\mathcal{A}_i\,\mathrm{d}x^i}$, here $\mathrm{d}_{\text{dR}}$~is the de~Rham differential on the Minkowski space\/-\/time. 
The conventional plus sign in the transcript of the Bianchi identities yields the minus sign in front of the adjoint operator $\mathrm{d}_{\text{dR}}^\dagger=-\mathrm{d}_{\text{dR}}$ in the gauge symmetries. Because the coefficients of the de~Rham differential, which spreads along the diagonal of the $(4\times4)$-\/matrix operator~$A$, are constant, the bi\/-\/differential Christoffel symbols~$\Gamma^A_{AA}$ vanish in this model.

It is remarkable that the Yang\/--\/Mills theories with non\/-\/abelian gauge groups
still do constitute a regular class of examples with the field\/-\/independent (and at most constant, whenever nonzero) coefficients in the bi\/-\/differential Christoffel symbols~$\Gamma^k_{ij}(\cdot,\cdot)$. On the other hand, 
\emph{gravity} produces the drastically more involved structure of these symbols with the explicit dependence on the unknown fields in their coefficients~\cite{FradkinVilkovisky1975}. 
This is immanent also to the Liouville\/-\/type systems (see section~\ref{SecToda}).
\end{example}

Let the operators~$A_1$,\ $\ldots$,\ $A_N\colon\widehat{P_1}\to\text{sym}\,\mathcal{E}\subset\varkappa(\pi)$ be the entire collection of the gauge symmetry generators
for the system~$\mathcal{E}$. For the sake of clarity, we 
always assume the off\/-\/shell 
validity of the Berends\/--\/Burgers\/--\/van~Dam hypothesis~\cite{BBvD1985} about the collective commutation closure for the images of these operators, see~\eref{Involutive}. Then the odd evolutionary vector field~\eref{Q} is the Becchi\/--\/Rouet\/--\/Stora\/--\/Tyutin differential~\cite{BRST}; in the model at hand, \emph{all} parity\/-\/reversed arguments~$\boldsymbol{b}_i$ belong to one module~$\Pi\widehat{P_1}$ and are known as the \emph{ghosts}, usually denoted by~$\boldsymbol{\gamma}_i$. The variational Lie algebroid encoded by the evolutionary differential~$\boldsymbol{Q}$ was named the \emph{gauge algebroid} in~\cite{BarnichBialowieza}.

As in the variational Poisson case (see section~\ref{SecHam}), we arrange the variables in pairs of opposite parity: even fields~$\boldsymbol{q}$ 
$\longleftrightarrow$ odd \emph{antifields}~$\boldsymbol{q}^\dagger$,
and odd ghosts~$\boldsymbol{\gamma}_i$ $\longleftrightarrow$ 
even \emph{antighosts}~$\boldsymbol{\gamma}_i^\dagger$.

The evolutionary BRST\/-\/field~$\boldsymbol{Q}$ lifts to the odd derivation~$\widetilde{\boldsymbol{Q}}$ on the horizontal infinite jet superspace that now contains
the antifield coordinates~$\boldsymbol{q}^\dagger_\tau$
and similarly, the antighosts $\boldsymbol{\gamma}_{i;\tau}^\dagger$, here~$|\tau|\geq0$. The induced velocities of the newly incorporated fibre variables are calculated in agreement with their geometric nature. 

First, we recall that the off\/-\/shell determining condition $\partial^{(\boldsymbol{q})}_\varphi(\boldsymbol{F})=\nabla(\boldsymbol{F})$ for $\varphi$~to be an infinitesimal symmetry of the equation~$\boldsymbol{F}=0$ states the existence of the linear total differential operator~$\nabla$ such that the right\/-\/hand side of the 
condition vanishes on\/-\/shell (i.e., by virtue of the equation and its differential consequences). Let us remember also that  the operator~$\nabla=\nabla_\varphi$ depends linearly on~$\varphi$ because the entire equality does~so.
Specifically for the gauge 
geometry at hand, this means that
$\partial^{(\boldsymbol{q})}_{A(\boldsymbol{\gamma})}(\boldsymbol{F})=\nabla_{A(\boldsymbol{\gamma})}(\boldsymbol{F})$ or, after the parity reversion $\Pi\colon\boldsymbol{F}
\rightleftarrows\boldsymbol{F}^\dagger=\boldsymbol{q}^\dagger$
of the linear entry~$\boldsymbol{F}$ in both sides of the
equality, we obtain the odd\/-\/parity 
component $\partial^{(\boldsymbol{q}^\dagger)}_{\textstyle \nabla_{A(\boldsymbol{\gamma})}(\boldsymbol{q}^\dagger)}$ of the lifting~$\widetilde{\boldsymbol{Q}}$ for~$\boldsymbol{Q}$.

Second, we see that --due to the chain rule-- the odd velocity of a Noether relation~$\Phi[\boldsymbol{F}]\in P_1$, or of the even antighost~$\boldsymbol{\gamma}^\dagger$
associated with it, is equal to
\[
\Bigl(\partial^{(\boldsymbol{q})}_{A(\boldsymbol{\gamma})} (A^\dagger)\Bigr) (\boldsymbol{F}) 
+ A^\dagger\bigl(\nabla_{A(\boldsymbol{\gamma})}
(\boldsymbol{F})\bigr).
\]
\begin{cor}  
The proper antifield\/-\/antighost lifting~$\widetilde{\boldsymbol{Q}}$ of the BRST\/-\/differential~$\boldsymbol{Q}$ reads
\[
\widetilde{\boldsymbol{Q}}=
 \partial^{(\boldsymbol{q})}_{\sum\limits_i 
  A_i(\boldsymbol{\gamma}_i)}
-\frac{1}{2}\sum\limits_k 
 \partial^{(\boldsymbol{\gamma}_k)}_{\sum\limits_{i,j}
  \Gamma^k_{ij}(\boldsymbol{\gamma}_i,\boldsymbol{\gamma}_j)}
+\partial^{(\boldsymbol{q}^\dagger)}_{\sum\limits_i \nabla_{A_i(\boldsymbol{\gamma}_i)}(\boldsymbol{q}^\dagger)}
+\sum\limits_k \partial^{(\boldsymbol{\gamma}_k^\dagger)}_{
 \textstyle \sum\limits_i\left(
  \partial^{(\boldsymbol{q})}_{A_i(\boldsymbol{\gamma}_i)} (A_k^\dagger) (\smash{\frac{\delta S}{\delta\boldsymbol{q}}}) 
   + A_k^\dagger\left(\nabla_{A_i(\boldsymbol{\gamma}_i)}
(\smash{\frac{\delta S}{\delta\boldsymbol{q}}})\right) \right)}.
\]
\end{cor}

The full Batalin\/--\/Vilkovisky
differential~$\boldsymbol{D}=\mathrm{d}_{\text{K\/-\/T}}+\widetilde{\boldsymbol{Q}}+\dots$ merges
the parity\/-\/reversing (and besides, vanishing on\/-\/shell)
Koszul\/--\/Tate differential,
\[
\mathrm{d}_{\text{K\/-\/T}}=
 \partial^{(\boldsymbol{q}^\dagger)}_{\textstyle
 \delta S/\delta\boldsymbol{q}}
+\sum\limits_k \partial^{(\boldsymbol{\gamma}_k^\dagger)}_{A_k^\dagger(\boldsymbol{q}^\dagger)},
\]
with the lifting~$\widetilde{\boldsymbol{Q}}$ of~$\boldsymbol{Q}$; however, $\boldsymbol{D}$~may 
contain further terms of the correction that ensures the off\/-\/shell equality~$\boldsymbol{D}^2=0$.

Let us now look at the problem of construction of the BV\/-\/differential from another perspective. 
The pairwise arrangement $\boldsymbol{q}^\dagger\leftrightarrow\boldsymbol{q}$,\ $\boldsymbol{\gamma}\leftrightarrow\boldsymbol{\gamma}^\dagger$ for the 
$\langle\,,\,\rangle$-\/dual variables of opposite parity
prescribes the composition of the variational Schouten bracket
$[\![\boldsymbol{\omega}_1,\boldsymbol{\omega}_2]\!]=
\bigl\langle\overrightarrow{\delta}\boldsymbol{\omega}_1\wedge
\overleftarrow{\delta}\boldsymbol{\omega}_2\bigr\rangle
$ (\cite{ZinnJustin1975} and, e.g., \cite{GomisParisSamuel,Antibracket}), 
now extending to all generations of the ghost\/-\/antighost
pairs:~$[\![\boldsymbol{\omega}_1,\boldsymbol{\omega}_2]\!]$
\begin{multline*}
{}=
\Bigl(\frac{\overrightarrow{\delta}\boldsymbol{\omega}_1}{\delta\boldsymbol{q}^\dagger}\langle\delta\boldsymbol{q}^\dagger,\delta\boldsymbol{q}\rangle\frac{\overleftarrow{\delta}\boldsymbol{\omega}_2}{\delta\boldsymbol{q}} 
-
\frac{\overrightarrow{\delta}\boldsymbol{\omega}_1}{\delta\boldsymbol{q}}\langle\delta\boldsymbol{q}^\dagger,\delta\boldsymbol{q}\rangle\frac{\overleftarrow{\delta}\boldsymbol{\omega}_2}{\delta\boldsymbol{q}^\dagger} \Bigr) 
+
\Bigl(\frac{\overrightarrow{\delta}\boldsymbol{\omega}_1}{\delta\boldsymbol{\gamma}}\langle\delta\boldsymbol{\gamma},
\delta\boldsymbol{\gamma}^\dagger\rangle\frac{\overleftarrow{\delta}\boldsymbol{\omega}_2}{\delta\boldsymbol{\gamma}^\dagger} 
-
\frac{\overrightarrow{\delta}\boldsymbol{\omega}_1}{\delta\boldsymbol{\gamma}^\dagger}\langle\delta\boldsymbol{\gamma},
\delta\boldsymbol{\gamma}^\dagger\rangle\frac{\overleftarrow{\delta}\boldsymbol{\omega}_2}{\delta\boldsymbol{\gamma}}\Bigr)
\\
{}=
\left[\frac{\overrightarrow{\delta}\boldsymbol{\omega}_1}{\delta\boldsymbol{q}^\dagger}\cdot
\frac{\overleftarrow{\delta}\boldsymbol{\omega}_2}{\delta\boldsymbol{q}}
-
\frac{\overrightarrow{\delta}\boldsymbol{\omega}_1}{\delta\boldsymbol{q}}\cdot
\frac{\overleftarrow{\delta}\boldsymbol{\omega}_2}{\delta\boldsymbol{q}^\dagger}\right]
+
\left[\frac{\overrightarrow{\delta}\boldsymbol{\omega}_1}{\delta\boldsymbol{\gamma}}\cdot
\frac{\overleftarrow{\delta}\boldsymbol{\omega}_2}{\delta\boldsymbol{\gamma}^\dagger}
-
\frac{\overrightarrow{\delta}\boldsymbol{\omega}_1}{\delta\boldsymbol{\gamma}^\dagger}\cdot
\frac{\overleftarrow{\delta}\boldsymbol{\omega}_2}{\delta\boldsymbol{\gamma}}
\right].
\end{multline*}
In this notation,\footnote{Likewise, the odd Laplacian is
\[
\Delta_{\text{BV}}=
\frac{\overrightarrow{\delta}}{\delta\boldsymbol{q}^\dagger}\circ
\frac{\overleftarrow{\delta}}{\delta\boldsymbol{q}} +
\frac{\overrightarrow{\delta}}{\delta\boldsymbol{\gamma}}
\circ
\frac{\overleftarrow{\delta}}{\delta\boldsymbol{\gamma}^\dagger}.
\]}
the BV\/-\/differential~$\boldsymbol{D}$ is
determined (\cite{ZinnJustin1975,Antibracket}, see also section~\ref{SecHam}) by the variational Schouten bracket of the BV\/-\/action~$\bar{\boldsymbol{S}}$ and the respective super\/-\/jet variables,
\[
\boldsymbol{D}=\partial^{(\boldsymbol{q})}_{\textstyle[\![\bar{\boldsymbol{S}},\boldsymbol{q}]\!]}+
\partial^{(\boldsymbol{q}^\dagger)}_{\textstyle[\![
\bar{\boldsymbol{S}},\boldsymbol{q}^\dagger]\!]}+
\partial^{(\boldsymbol{\gamma})}_{\textstyle[\![
\bar{\boldsymbol{S}},\boldsymbol{\gamma}]\!]}+
\partial^{(\boldsymbol{\gamma}^\dagger)}_{\textstyle[\![\bar{\boldsymbol{S}},\boldsymbol{\gamma}^\dagger]\!]},
\qquad \text{i.e.,}\quad
\boldsymbol{D}(\Xi)=[\![\bar{\boldsymbol{S}},\Xi]\!],
\]
where the extended action functional 
$\bar{\boldsymbol{S}}=-\boldsymbol{S}$
has even parity and~$\boldsymbol{S}$
equals~\cite{DuboisViolette,HenneauxCharge,
KostantSternberg}, with the conventional
Einstein's summation over repeated indices
and Dirac's ordering of (co)vectors,
\[
\boldsymbol{S}=S+    
\langle\boldsymbol{q}^\dagger,A_k(\boldsymbol{\gamma}_k)\rangle
-\frac{1}{2}\left<\Gamma_{ij}^k(\boldsymbol{\gamma}_i,
\boldsymbol{\gamma}_j),\boldsymbol{\gamma}^\dagger_k\right>+
\left<\text{correction terms}\right>.
\]
The possible necessity to introduce the correction terms of higher polynomial orders in $\boldsymbol{\gamma}$,\ $\boldsymbol{\gamma}^\dagger$, or $\boldsymbol{q}^\dagger$
is legitimate, e.g., when the coefficients of the bi\/-\/differential symbols~$\Gamma_{ij}^k$ depend explicitly on the fields~$\boldsymbol{q}$ and hence produce the redundant terms in the velocity of~$\boldsymbol{q}^\dagger$, which thus must be cancelled out properly. 
The renowned paper~\cite{CattaneoFelder2000} (c.f.~\cite{FulpLadaStasheffSrni}) contains a perfect example of a gauge model with one generator~$A$ of its gauge symmetries and an explicit calculation of the (zero\/-\/order) bi\/-\/differential Christoffel symbols~$\Gamma^A_{AA}$, BRST-{} and BV\/-\/differentials, and the master\/-\/action~$\boldsymbol{S}$, see~\cite[p.~595]{CattaneoFelder2000} where the parity reversion~$\Pi$ is the `promotion' and the 
BRST-\/field~$
{\boldsymbol{Q}}$ is denoted by~$\delta_0$.

The cohomological condition~$[\boldsymbol{D},\boldsymbol{D}]=0$ is equivalent to the classical master 
equation $[\![\boldsymbol{S},\boldsymbol{S}]\!]=0$.
The BV\/-\/cohomology with respect to the differential~$\boldsymbol{D}$ (or $[\![\boldsymbol{S},\cdot]\!]$) provides the resolvent for the algebra~$C^\infty(\mathcal{E})$ of \emph{observables} in the model and opens a way for the quantization of gauge systems.

\begin{prb}
Describe the variational Poisson structures that are
pertinent to gauge\/-\/invariant systems.
\end{prb}

The most recent techniques of~\cite{GolovkoDisser,JK2011WDVV,JKspt2011} allow --or soon will allow-- the regular search for variational Poisson structures on non\/-\/evolutionary gauge\/-\/invariant models by using the respresentation of the structures via~$\boldsymbol{Q}^2=0$. It must be noted however,
that the gauge algebroids are endowed \textit{a priori} with the evolutionary differentials which suit well for the quantization of the systems. Therefore, the variational Poisson structures --whenever found for a given gauge model-- will either provide only a few of its symmetries or, if there is a functional freedom in the domains of the Hamiltonian operators, will definitely \emph{not} provide the entire Lie algebra of infinitesimal gauge transformations because of the apparent difference between the domains of definition for the Hamiltonian operators and ordinary gauge generators (compare the diagrams in this section and in section~\ref{SecHam}).
This confirms 
that the bi\/-\/Hamiltonian \emph{complete integrability} paradigm (\cite{Magri}, see~\cite{JKspt2011}), which relies on the construction of the Poisson pencils for hierarchies of evolutionary systems, becomes insufficient in the geometry of gauge fields.

\newpage
\section*{Conclusion}
The realization of nonlinear systems, or parts of the information which they carry, in terms of homological evolutionary vector fields~$\boldsymbol{Q}$ is not specific to gauge\/-\/invariant models only. Indeed, other natural classes of partial differential equations can be addressed with convenience from the same viewpoint. Moreover, even for the already known gauge systems the available approach is ready to capture more of their geometry without any extra modifications.

For a long time, 
the physical and mathematical incarnations 
of the  
notion of Lie algebroids over infinite jet spaces developed in parallel, the progress on the math side being seriously retarded with respect to the demands from physics.
Seemingly, the 
exchange between the two theories, foresighted in~\cite{BelavinSbornik}, channeled through 
the construction of the \emph{variational} Schouten bracket.
Nowadays, the systematic and well\/-\/motivated calculation of the variational Poisson cohomology only begins~--- while the BRST-{} and BV\/-\/cohomology theories are an established domain of research and the subject of a vastest literature.
In this review we traced the links between the variational Poisson algebroids, gauge algebroids, and --playing the r\^ole of 
mediators-- 2D~Toda\/-\/like systems also viewed as variational Lie algebroids. 

\ack
The author is grateful to the Organizing committee
of 7th~International Workshop
`Quantum Theory and Symmetries\/--\/7' 
(August 7--13, 2011; CVUT Prague, Czech Republic)
for a welcome and warm atmosphere during the meeting.
The author thanks G.~Barnich, B.~A.~Dubrovin,
G.~Felder, E.~A.~Ivanov,
S.~O.~Krivonos, 
J.~W.~van de~Leur, 
M.~Schli\-chen\-ma\-ier, M.~A.~Semenov\/--\/Tian\/-\/Shansky, 
and M.~A.~Vasiliev for helpful discussions and 
stimulating remarks.
Also, the author thanks J.~Kra\-sil'\-shchik and A.~Verbovetsky    
for constructive criticisms and interesting correspondence.

This research 
was supported in part by NWO~VENI 
grant~639.031.623 (Utrecht) and JBI~RUG 
project~103511 (Groningen). A~part of this research was done while the author 
was visiting at the $\smash{\text{IH\'ES}}$
(Bures\/-\/sur\/-\/Yvette); the financial support and hospitality of this institution are gratefully acknowledged.

\section*{Appendix: The hunting of the variational Lie algebroids}
We now list several operators which we obtained in a fixed system of
local coordinates and whose images are then closed under commutation.
We denote by~$u(x)$ the field and, for brevity, we use the notation~$\{\!\{p,q\}\!\}_A$ instead of~$\Gamma^A_{AA}(p,q)$.
Let us fix the weights $|u|=2$, $|\mathrm{d}/\mathrm{d}x|=1$ that originate from the scaling invariance of the KdV equation $u_t=-\tfrac{1}{2}u_{xxx}+3uu_x$;
we have that $|\mathrm{d}/\mathrm{d}t|=3$.
Using the method of undetermined coefficients, we performed the search for scalar 
operators that satisfy~\eref{DefFrob} and which
are homogeneous with respect to the weights not
greater than~$7$. We obtained two compatible Hamiltonian operators
$A_1^{\text{KdV}}=\mathrm{d}/\mathrm{d} x$ and 
$A_2^{\text{KdV}}=-\tfrac{1}{2}(\mathrm{d}/\mathrm{d} x)^3+2u\,\mathrm{d}/\mathrm{d} x+u_x$,
the odd powers $(\mathrm{d}/\mathrm{d} x)^{2n+1}$ of~$\mathrm{d}/\mathrm{d} x$, 
and the Hamiltonian operator 
\[ 
u^2\,\left(\tfrac{\mathrm{d}}{\mathrm{d} x}\right)^3
 +3uu_x\,\left(\tfrac{\mathrm{d}}{\mathrm{d} x}\right)^2
 +3uu_{xx}\,\tfrac{\mathrm{d}}{\mathrm{d} x} + uu_{xxx}.
\]
Also, there are four non\/-\/skew\/-\/adjoint operators 
with involutive images,
\begin{align*}
A_4^{(6)}&=u^3-u_x^2,\qquad A_5^{(6)}=2u_x^2-uu_{xx}
  -2uu_x\,\tfrac{\mathrm{d}}{\mathrm{d} x}+u^2\,\left(\tfrac{\mathrm{d}}{\mathrm{d} x}\right)^2,\\
  \{\!\{p,q\}\!\}_{A_4^{(6)}}&=2u_x\cdot (p q_x-p_x q),\qquad
\{\!\{p,q\}\!\}_{A_5^{(6)}}=-2u_x\cdot (p q_x-p_x q)+u\cdot (p q_{xx}-p_{xx} q);
\\  
A_8^{(7)}&=u_x^2\,\tfrac{\mathrm{d}}{\mathrm{d} x}-2uu_{xx}\,\tfrac{\mathrm{d}}{\mathrm{d} x}
 -4uu_x\left(\tfrac{\mathrm{d}}{\mathrm{d} x}\right)^2
 -4u^2\,\left(\tfrac{\mathrm{d}}{\mathrm{d} x}\right)^3,\qquad
  \{\!\{p,q\}\!\}_{A_8^{(7)}}=u^2\cdot (p q_x-p_x q);\\
A_9^{(7)}&=-2u_xu_{xx}-u_x^2\,\tfrac{\mathrm{d}}{\mathrm{d} x},\qquad
\{\!\{p,q\}\!\}_{A_9^{(7)}}=8u_{xx}\cdot (p q_x-p_x q)+
2u_x\cdot (p q_{xx}-p_{xx} q).
\end{align*}
Finally, we have found the operators that contain arbitrary functions:
$f(u)(\mathrm{d}/\mathrm{d} x)^n$ and $f(u)u^2$ with vanishing 
Christoffel symbols~$\{\!\{\,,\,\}\!\}_A$, and also
\begin{align*}
A_3&=f(u)u_x,&
   \{\!\{p,q\}\!\}_{A_3}&=f(u)\bigl(p_xq-pq_x\bigr);
\\
A_4&=f(u)\bigl(u\,\tfrac{\mathrm{d}}{\mathrm{d} x}-u_x\bigr),&
   \{\!\{p,q\}\!\}_{A_4}&=f(u)\bigl(p q_x-p_x q\bigr).
\end{align*}
The transformation rules for the sections that constitute
the domains of the operators 
$A_3$,\ $A_4$,\ $A_4^{(6)}$,\ $\ldots$,\ $A_9^{(7)}$
are yet to be found.  
Likewise, the interpretation as variational anchors and the understanding of the physics of the models remain an open problem for the infinite class of linear total differential operators with involutive images which were found in~\cite{Sanders,SokolovUMN}.  

\section*{References}   

\end{document}